\documentclass[a4paper,12pt]{article}
\usepackage{amssymb}
\usepackage{amsmath}
\usepackage{bm}
\usepackage{boxedminipage}
\usepackage[pdftex]{graphicx}
\usepackage{graphicx}

\@addtoreset{equation}{section} \makeatother

\newif\ifpdf \ifx\pdfoutput\undefined \pdffalse
\else \pdfoutput=1 \pdftrue \fi \ifpdf \else \fi
\input epsf

\begin{document}

\ifpdf\DeclareGraphicsExtensions{.pdf, .jpg, .tif} \else%
\DeclareGraphicsExtensions{.eps, .jpg} \fi
\begin{titlepage}

    \thispagestyle{empty}
    \begin{flushright}
        \hfill{CERN-PH-TH/2008-020}\\
    \end{flushright}

    \vspace{10pt}
    \begin{center}
        { \Large{\textbf{$d=4$ Black Hole Attractors in $N=2$ Supergravity} \\\vspace{12pt}
 \textbf{with Fayet-Iliopoulos Terms}}}

        \vspace{18pt}

        {\bf Stefano Bellucci$^{\spadesuit}$, \ Sergio Ferrara$^{\diamondsuit,\spadesuit}$, \ Alessio Marrani$^{\heartsuit,\spadesuit}$ and \ Armen Yeranyan$^{\spadesuit,\clubsuit}$}

        \vspace{20pt}

        {$\spadesuit$ \it INFN - Laboratori Nazionali di Frascati, \\
        Via Enrico Fermi 40,00044 Frascati, Italy\\
        \texttt{bellucci,marrani,ayeran@lnf.infn.it}}

        \vspace{10pt}

        {$\diamondsuit$ \it Physics Department,Theory Unit, CERN, \\
        CH 1211, Geneva 23, Switzerland\\
        \texttt{sergio.ferrara@cern.ch}}

        \vspace{10pt}

        {$\heartsuit$ \it Museo Storico della Fisica e\\
        Centro Studi e Ricerche ``Enrico Fermi"\\
        Via Panisperna 89A, 00184 Roma, Italy}

        \vspace{10pt}

        {$\clubsuit$ \it Department of Physics, Yerevan State University, \\Alex  Manoogian St., 1, Yerevan,
        375025, Armenia}

        \vspace{5pt}

        \vspace{40pt}

        {ABSTRACT}
    \end{center}

We generalize the description of the $d=4$ \textit{Attractor
Mechanism} based on an effective black hole (BH) potential to the
presence of a gauging which does not modify the derivatives of the
scalars and does not involve hypermultiplets. The obtained results
do not rely necessarily on supersymmetry, and they can be extended
to $d>4$, as well.

Thence, we work out the example of the $stu$ model of $N=2$
supergravity in the presence of Fayet-Iliopoulos terms, for the
supergravity analogues of the \textit{magnetic} and $\mathit{%
D0-D6}$ BH charge configurations, and in three different symplectic
frames: the $\left( SO\left( 1,1\right) \right) ^{2}$, $SO\left(
2,2\right) $ covariant and $SO\left( 8\right) $-truncated ones.

The attractive nature of the critical points, related to the
semi-positive definiteness of the Hessian matrix, is also studied.


\end{titlepage}
\newpage\tableofcontents

\section{\label{Intro}Introduction}

The so-called \textit{Attractor Mechanism} was discovered in the mid 90's
\cite{FKS}-\nocite{Strom,FK1,FK2}\cite{FGK} in the context of extremal BPS
(Bogomol'ny-Prasad-Sommerfeld) black holes (BHs). Recently, extremal BH
\textit{attractors} have been intensively studied \cite{Sen-old1}--\nocite
{GIJT,Sen-old2,K1,TT,G,GJMT,Ebra1,K2,Ira1,Tom,
BFM,AoB-book,FKlast,Ebra2,BFGM1,rotating-attr,K3,Misra1,Lust2,Morales,BFMY,
CdWMa,
DFT07-1,BFM-SIGRAV06,Cer-Dal,ADFT-2,Saraikin-Vafa-1,Ferrara-Marrani-1, TT2,
ADOT-1,fm07,CCDOP,Misra2,Astefanesei,Anber,Myung1,Ceresole,BMOS-1,Hotta,
Gao,
PASCOS07,Sen-review,Belhaj1,AFMT1,Gaiotto1,BFMS1,GLS1,ANYY1,review-Kallosh,Cai-Pang}
\cite{Vaula}, mainly due to the (re)discovery of new classes of scalar
attractor configurations. Differently from the BPS ones, such configurations
do not saturate the BPS bound \cite{BPS} and, when considering a
supergravity theory, they break all supersymmetries at the BH event horizon.

The \textit{Attractor Mechanism} was firstly discovered in \textit{ungauged}
supergravities, with various supercharges and in various space-time
dimensions. In all cases such a mechanism of purely charge-dependent
stabilization of the scalars at the event horizon of an extremal BH can be
\textit{critically implemented}, \textit{i.e.} the \textit{attractor}
configurations are nothing but the critical points of a suitably defined,
positive definite \textit{BH effective potential} function $V_{BH}$. The
formalism based on $V_{BH}$ is intrinsically \textit{off-shell}, \textit{i.e.%
} it holds in the whole scalar manifold, and it allows one to determine the
classical Bekenstein-Hawking \cite{BH1} BH entropy by going \textit{on-shell}%
, that is by evaluating $V_{BH}$ at its \textit{non-degenerate} critical
points, \textit{i.e.} at those critical points for which $V_{BH}\neq 0$.
Furthermore, one can study the stability of the attractors by considering
the sign of the eigenvalues of the (real form of the) Hessian matrix of $%
V_{BH}$ at its critical points. Actually, only those critical points
corresponding to \textit{all} strictly positive Hessian eigenvalues should
be rigorously named \textit{attractors}.

In this work we deal with the extension of the \textit{effective potential
formalism} in the presence of a particularly simple \textit{gauging}. In
other words, we determine the generalization of $V_{BH}$ to an \textit{%
effective potential} $V_{eff}$, which takes into account also the (not
necessarily positive definite) potential $V$ originated from the gauging. We
show that the \textit{Attractor Mechanism} can still be implemented in a
critical fashion, \textit{i.e.} that the stabilized configurations of the
scalar fields, depending on the BH conserved electric and magnetic charges
and on the gauge coupling constant $g$, can be determined as the \textit{%
non-degenerate} critical points of $V_{eff}$. Concerning the $d=4$ metric
background, we consider a dyonic, extremal, static, spherically symmetric
BH, which generally is asymptotically \textit{non-flat}, due to the presence
of a non-vanishing gauge potential $V$. As we will discuss further below,
such a space-time background is \textit{non-supersymmetric}, and thus
\textit{all} attractors in such a framework will break \textit{all}
supersymmetries down.

As a particular working example, next we consider the so-called
$stu$ model with Fayet-Iliopoulos (FI) terms, in two manageable BH
charge configurations, namely in the supergravity analogues of the
\textit{magnetic} and $\mathit{D0-D6}$ ones. It is worth
anticipating here that the obtained results (and their physical
meaning) crucially depend on the choice of the symplectic basis used
to describe the physical system at hand. Such a feature strongly
distinguishes the gauged case from the ungauged one, the latter
being insensitive to the choice of the symplectic basis. However,
such a phenomenon, observed some time ago \textit{e.g.} in
\cite{Freedman-Schwarz}, can be easily understood by noticing that
in general the gauge potential $V$ is independent of the BH charges,
and thus it is necessarily not symplectic invariant. Indeed, as we
explicitly compute, the resulting explicit form of the
\textit{generalized} BH effective potential $V_{eff}$, its critical
points, their positions in the scalar manifold, their stability
features and the corresponding BH entropies generally do depend on
the symplectic frame being considered.

It is here worth noticing that the $SO\left( 8\right) $-truncated
symplectic basis (treated in Subsect. \ref{Duff-Liu}) has the
remarkable
property that the FI potential $V_{FI,SO\left( 8\right) }$ (given by Eq. (%
\ref{admor}) below) admits critical points at a finite distance from
the origin. This is not the case for the FI potentials in the
$\left( SO\left( 1,1\right) \right) ^{2}$ and $SO\left( 2,2\right) $
covariant bases treated in Subsects. \ref{stu} and \ref{Heterotic}.
Indeed, $V_{FI,\left( SO\left( 1,1\right) \right) ^{2}}$ and
$V_{FI,SO\left(
2,2\right) }$ (respectively given by Eqs. (\ref{scalpotstu}) and (\ref{dom1}%
) below) only have a runaway behavior, admitting only minima at an infinite
distance from the origin.

Finally, we address the issue of the stability of the various
obtained \textit{gauged attractors}. in light of the previous
statement, not surprisingly we find that the lifting features of the
two non-BPS $Z\neq 0$ \textit{massless} Hessian modes pertaining to
the ungauged limit ($g=0$) of the $stu$ model depend on the
symplectic basis being considered.\medskip

The plan of the paper is as follows.

In Sect. \ref{General} we exploit a general treatment of the \textit{%
effective potential formalism} in the presence of a gauging which does not
modify the derivatives of the scalars and does not involve hypermultiplets;
such a treatment does not rely necessarily on supersymmetry\footnote{%
In some respect, a similar treatment in $d=5$ was recently given in
\cite {ANYY1}.}. Then, in Sect. \ref{General-stu} we apply the
obtained general results to the case of the $stu$ model in the
presence of Fayet-Iliopoulos terms, in three different symplectic
bases: the so-called $\left( SO\left( 1,1\right) \right) ^{2}$,
$SO\left( 2,2\right) $ covariant and $SO\left( 8\right) $-truncated
ones, respectively considered in Subsects. \ref{stu},
\ref{Heterotic} and \ref{Duff-Liu}. Thus, in Sect.
\ref{BH-Entropies} we compute, for the so-called magnetic and
$D0-D6$ BH charge configurations (respectively treated in Subsects.
\ref{Magnetic} and \ref{D2-D6}), the critical points of the
effective potential $V_{eff}$ of the $stu$ model in the symplectic
frames introduced above, and then compare the resulting BH
entropies, obtaining that the results will generally depend on the
symplectic frame considered. Sect. \ref {Stability} is devoted to
the analysis of the stability of the critical points of $V_{eff}$
computed in Sect. \ref{BH-Entropies}, in both the magnetic and
$D0-D6$ BH charge configurations (respectively treated in Subsects.
\ref{Magnetic2} and \ref{D0-D6-2}). Final comments and ideas for
further developments are given in Sect. \ref{Conclusion}.

\section{\label{General}General Analysis}

Let us start from the bosonic sector of \textit{gauged} $N=2$, $d=4$
supergravity. We consider a gauging which does not modify the derivatives of
the scalars and does not involve hypermultiplets. The appropriate action has
the following form ($\mu =01,2,3$, $a=1,...,n_{V}$ and $\Lambda
=0,1,....,n_{V}$, $n_{V}$ being the number of vector multiplets; see \textit{%
e.g.} \cite{N=2-big})
\begin{equation}
S=\int \left( -\frac{1}{2}R+G_{a\bar{b}}(z,\bar{z})\partial _{\mu
}z^{a}\partial ^{\mu }\bar{z}^{\overline{b}}+\mu _{\Lambda \Sigma }(z,\bar{z}%
)\mathcal{F}_{\mu \nu }^{\Lambda }\mathcal{F}^{\Sigma |\,\mu \nu }+\nu
_{\Lambda \Sigma }(z,\bar{z})\mathcal{F}_{\mu \nu }^{\Lambda }{^{\ast }%
\mathcal{F}}^{\Sigma |\,\mu \nu }-V(z,\bar{z})\right) \sqrt{-g}d^{4}x^{\mu },
\label{action}
\end{equation}
where $\mu _{\Lambda \Sigma }=Im\mathcal{N}_{\Lambda \Sigma }$, $\nu
_{\Lambda \Sigma }=Re\mathcal{N}_{\Lambda \Sigma }$, $\mathcal{F}_{\mu \nu
}^{\Lambda }=\frac{1}{2}(\partial _{\mu }A_{\nu }^{\Lambda }-\partial _{\nu
}A_{\mu }^{\Lambda })$ and $^{\ast }$ denotes Hodge duality ($\varepsilon
^{0123}=-\varepsilon _{0123}\equiv -1$)
\begin{equation}
{^{\ast }\mathcal{F}}^{\Lambda |\,\mu \nu }=\frac{1}{2\sqrt{-g}}\varepsilon
^{\mu \nu \lambda \rho }\mathcal{F}_{\lambda \rho }^{\Lambda }.
\end{equation}

Varying the action (\ref{action}) with respect to $g^{\mu \nu }$, $z^{a}$
and $A_{\mu }^{\Lambda }$ one can easily find the following Eqs. of motion:
\begin{eqnarray}
&&R_{\mu \nu }-\frac{1}{2}Rg_{\mu \nu }-2G_{a\bar{b}}\partial _{(\mu
}z^{a}\partial _{\nu )}\bar{z}^{\overline{b}}+G_{a\bar{b}}\partial _{\lambda
}z^{a}\partial ^{\lambda }\bar{z}^{\overline{b}}g_{\mu \nu }=T_{\mu \nu
}+Vg_{\mu \nu };  \label{grav} \\
&&\frac{G_{a\bar{b}}}{\sqrt{-g}}\partial _{\mu }(\sqrt{-g}g^{\mu \nu
}\partial _{\nu }\bar{z}^{\overline{b}})+\frac{\partial G_{a\bar{b}}}{%
\partial \bar{z}^{\overline{c}}}\partial _{\lambda }\bar{z}^{\overline{b}%
}\partial ^{\lambda }\bar{z}^{\overline{c}}=\frac{\partial \mu _{\Lambda
\Sigma }}{\partial z^{a}}\mathcal{F}_{\mu \nu }^{\Lambda }\mathcal{F}%
^{\Sigma |\,\mu \nu }+\frac{\partial \nu _{\Lambda \Sigma }}{\partial z^{a}}%
\mathcal{F}_{\mu \nu }^{\Lambda }{^{\ast }\mathcal{F}}^{\Sigma |\,\mu \nu }-%
\frac{\partial V}{\partial z^{a}};  \label{scal} \\
&&\varepsilon ^{\mu \nu \lambda \rho }\partial _{\nu }\mathcal{G}_{\Lambda
|\,\lambda \rho }=0,  \label{elek}
\end{eqnarray}
where $\mathcal{G}_{\Lambda |\,\mu \nu }\equiv \nu _{\Lambda \Sigma }%
\mathcal{F}_{\mu \nu }^{\Sigma }-\mu _{\Lambda \Sigma }{^{\ast }\mathcal{F}}%
_{\mu \nu }^{\Sigma }$ and
\begin{equation}
T_{\mu \nu }\equiv 4\mu _{\Lambda \Sigma }\mathcal{F}_{\mu \lambda
}^{\Lambda }\mathcal{F}_{\nu \rho }^{\Sigma }g^{\lambda \rho }-\mu _{\Lambda
\Sigma }\mathcal{F}_{\lambda \rho }^{\Lambda }\mathcal{F}^{\Sigma |\,\lambda
\rho }g_{\mu \nu }.  \label{elekten}
\end{equation}

Moreover, one also obtains
\begin{equation}
\varepsilon ^{\mu \nu \lambda \rho }\partial _{\nu }\mathcal{F}_{\lambda
\rho }^{\Lambda }=0,  \label{bi}
\end{equation}
which directly follows from the very definition of $\mathcal{F}_{\lambda
\rho }^{\Lambda }$, and it is nothing but the Bianchi identities.

Let us now consider the dyonic, static, spherically symmetric extremal black
hole (BH) background. For this case, the most general metric reads
\begin{equation}
ds^{2}=e^{2A(r)}dt^{2}-e^{2B(r)}dr^{2}-e^{2C(r)}r^{2}(d\theta ^{2}+\sin ^{2}{%
\theta }d\varphi ^{2}).  \label{met}
\end{equation}
It is worth pointing out that only two functions out of $A(r)$, $B(r)$ and $%
C(r)$ are independent, one of the three can be absorbed by a suitable
redefinition of the radial coordinate $r$. As a consequence of the staticity
and spherical symmetry, all scalar fields depend only of $r$, and
intensities of electrical and magnetic fields are directed along the radial
direction.

Let us now investigate the Eq. (\ref{elek}) of the electromagnetic field and
the Bianchi identities (\ref{bi}). The electromagnetic field strength has
only two non-zero components, namely $\mathcal{F}_{01}^{\Lambda }$ and $%
\mathcal{F}_{23}^{\Lambda }$, corresponding to radially directed electrical
and magnetic fields, respectively. Indeed, by solving the Bianchi identities
(\ref{bi}), one gets that $\mathcal{F}_{01}^{\Lambda }=\mathcal{F}%
_{01}^{\Lambda }(r)$ and $\mathcal{F}_{23}^{\Lambda }=\mathcal{F}%
_{23}^{\Lambda }(\theta )$. The same holds for $\mathcal{G}_{\Lambda |\,\mu
\nu }$, because by duality arguments Eq. (\ref{elek}) is nothing but the
Bianchi identities for $\mathcal{G}_{\Lambda |\,\mu \nu }$. Substituting $%
\mathcal{F}_{01}^{\Lambda }$ and $\mathcal{F}_{23}^{\Lambda }$ into $%
\mathcal{G}_{\Lambda |\,01}$ and $\mathcal{G}_{\Lambda |\,23}$ and recalling
the peculiar dependence on $r$ and $\theta $, one finally gets the following
solution for the non-vanishing components of the electromagnetic field
strength:
\begin{eqnarray}
&&\mathcal{F}_{01}^{\Lambda }=\frac{e^{A+B-2C}}{2r^{2}}(\mu )^{-1|\,\Lambda
\Sigma }(\nu _{\Sigma \Gamma }p^{\Gamma }-q_{\Sigma });  \notag \\
&&\mathcal{F}_{23}^{\Lambda }=-\frac{1}{2}p^{\Lambda }\sin {\theta },
\label{f}
\end{eqnarray}
where the constants $q_{\Lambda }$ and $p^{\Lambda }$ respectively are the
electrical and magnetic charges of the BH, \textit{i.e.} the fluxes
\begin{equation}
\left(
\begin{array}{c}
p^{\Lambda } \\
q_{\Lambda }
\end{array}
\right) \equiv \frac{1}{4\pi }\left(
\begin{array}{c}
\int_{S_{\infty }^{2}}\mathcal{F}^{\Lambda } \\
\int_{S_{\infty }^{2}}\mathcal{G}_{\Lambda }
\end{array}
\right) .
\end{equation}

$T_{\mu \nu }$ can be computed by substituting Eqs. (\ref{f}) into Eq. (\ref
{elekten}). By doing so, one obtains
\begin{equation}
T_{0}^{0}=T_{1}^{1}=-T_{2}^{2}=-T_{3}^{3}=\frac{e^{-4C}}{r^{4}}V_{BH},
\label{tv}
\end{equation}
where $V_{BH}$ is the standard, so-called \textit{BH potential}, defined as
(see \textit{e.g.} \cite{FGK})
\begin{equation}
V_{BH}\equiv -\frac{1}{2}(p^{\Lambda },\,q_{\Lambda })\left(
\begin{array}{cc}
\mu _{\Lambda \Sigma }+\nu _{\Lambda \Gamma }(\mu )^{-1|\,\Gamma \Pi }\nu
_{\Pi \Sigma } & \;-\nu _{\Lambda \Gamma }(\mu )^{-1|\,\Gamma \Sigma } \\
-(\mu )^{-1|\,\Lambda \Gamma }\nu _{\Gamma \Sigma } & \;(\mu )^{-1|\,\Lambda
\Sigma }
\end{array}
\right) \left(
\begin{array}{c}
p^{\Sigma } \\
q_{\Sigma }
\end{array}
\right) .  \label{VBH-def}
\end{equation}
One can also easily check that
\begin{equation}
\frac{\partial \mu _{\Lambda \Sigma }}{\partial z^{a}}\mathcal{F}_{\mu \nu
}^{\Lambda }\mathcal{F}^{\Sigma |\,\mu \nu }+\frac{\partial \nu _{\Lambda
\Sigma }}{\partial z^{a}}\mathcal{F}_{\mu \nu }^{\Lambda }{^{\ast }\mathcal{F%
}}^{\Sigma |\,\mu \nu }=-\frac{e^{-4C}}{r^{4}}\frac{\partial V_{BH}}{%
\partial z^{a}}.  \label{scaltv}
\end{equation}

Let us now write down Eqs. (\ref{grav}) and (\ref{scal}) explicitly. In
order to do this, we substitute the metric (\ref{met}) and Eqs. (\ref{tv})
and (\ref{scaltv}) into them, getting the following independent Eqs.:
\begin{equation}
\begin{array}{l}
-e^{-2B}\left[ C^{\prime }(C^{\prime }+2A^{\prime })+\frac{2}{r}(C^{\prime
}+A^{\prime })+\frac{1}{r^{2}}\left( 1-e^{2(B-C)}\right) \right]
+e^{-2B}z^{a\,\prime }\bar{z}^{\overline{b}\,\prime }G_{a\bar{b}}=\frac{%
e^{-4C}}{r^{4}}V_{BH}+V; \\
\\
-e^{-2B}\left[ A^{\prime \prime }+C^{\prime \prime }+A^{\prime }(A^{\prime
}-B^{\prime })+C^{\prime }(C^{\prime }-B^{\prime }+A^{\prime })+\frac{2}{r}%
(A^{\prime }-B^{\prime }+2C^{\prime })\right] + \\
\\
-e^{-2B}z^{a\,\prime }\bar{z}^{\overline{b}\,\prime }G_{a\bar{b}}=-\frac{%
e^{-4C}}{r^{4}}V_{BH}+V; \\
\\
e^{-2B}\left[ \bar{z}^{\overline{b}\,\prime \prime }G_{a\bar{b}}+\bar{z}^{%
\overline{b}\,\prime }\bar{z}^{\overline{c}\,\prime }G_{a\bar{b},\,\bar{c}}+%
\bar{z}^{\overline{b}\,\prime }G_{a\bar{b}}(A^{\prime }-B^{\prime
}+2C^{\prime }+\frac{2}{r})\right] =\frac{e^{-4C}}{r^{4}}\frac{\partial
V_{BH}}{\partial z^{a}}+\frac{\partial V}{\partial z^{a}},
\end{array}
\label{eq}
\end{equation}
where the prime denotes derivative with respect to $r$.

Let us now perform the \textit{near-horizon limit} of Eqs. (\ref{eq}).
Within the assumptions (\ref{met}) made for the 4-dimensional BH background,
extremality implies that the near-horizon geometry is $AdS_{2}\times S_{2}$.
Thus, in the \textit{near-horizon limit} the functions appearing in the
metric (\ref{met}) read
\begin{equation}
A=-B=\log {\frac{r}{r_{A}}},\quad C=\log {\frac{r_{H}}{r}},
\label{near-horizon}
\end{equation}
where the relation $A=-B$ has been obtained by suitably redefining the
radial coordinate, and the constants $r_{A}$, $r_{H}$ are the \textit{radii}
of $AdS_{2}$ and $S_{2}$, respectively.

By assuming that the scalar fields $z^{a}$ are regular when approaching the
BH event horizon ($\exists $ $\lim_{r\rightarrow r_{H}}$ $z^{a}\left(
r\right) \equiv z_{H}^{a}$, $\left| z_{H}^{a}\right| <\infty $), the \textit{%
near-horizon limit} of Eqs. (\ref{eq}) read
\begin{gather}
\frac{1}{r_{H}^{2}}=\frac{1}{r_{H}^{4}}V_{BH}+V;  \label{rh} \\
\frac{1}{r_{A}^{2}}=\frac{1}{r_{H}^{4}}V_{BH}-V;  \label{ra} \\
\frac{1}{r_{H}^{4}}\frac{\partial V_{BH}}{\partial z^{a}}+\frac{\partial V}{%
\partial z^{a}}=0,  \label{min}
\end{gather}
respectively yielding the following solutions:
\begin{gather}
r_{H}^{2}=V_{eff}\left( z_{H},\overline{z}_{H}\right) ;  \label{CERNN2} \\
r_{A}^{2}=\left. \frac{V_{eff}}{\sqrt{1-4V_{BH}|V|}}\right| _{z=z_{H}};
\label{CERNN3} \\
\left. \frac{\partial V_{eff}}{\partial z^{a}}\right| _{z=z_{H}}=0,
\label{crit-cond}
\end{gather}
where
\begin{equation}
V_{eff}\equiv \frac{1-\sqrt{1-4V_{BH}V}}{2V}  \label{Veff-def}
\end{equation}
and Eq. (\ref{crit-cond}) characterizes the horizon configurations $%
z_{H}^{a} $ of the scalars as the critical points of $V_{eff}$ in the scalar
manifold. In other words, near the BH horizon the scalars $z^{a}$ are
\textit{attracted} towards the purely charge-dependent configurations $%
z_{H}^{a}\left( p,q\right) $ satisfying Eq. (\ref{crit-cond}).

In the considered spherically symmetric framework the Bekenstein-Hawking
\cite{BH1}\textbf{\ }BH entropy reads
\begin{equation}
S_{BH}=\frac{A_{H}}{4}=\pi r_{H}^{2}=V_{eff}\left( z_{H},\overline{z}%
_{H}\right)
\end{equation}
Thus, in the presence of a non-vanishing scalar potential $V$ in the $d=4$
Lagrangian density, the \textit{Attractor Mechanism} \cite{FKS}-\nocite
{Strom,FK1,FK2}\cite{FGK} still works, with the usual $V_{BH}$ replaced by $%
V_{eff}$ defined by Eq. (\ref{Veff-def}).

From its very definition (\ref{Veff-def}), $V_{eff}$ is defined in the
region of the scalar manifold where $V_{BH}V\leqslant \frac{1}{4}$.
Moreover, analogously to $V$, it is not necessarily positive, as instead $%
V_{BH}$ is. It also holds that
\begin{gather}
\lim_{V\rightarrow 0}V_{eff}=V_{BH}; \\
\lim_{V_{BH}\rightarrow 0^{+}}V_{eff}=0.
\end{gather}
By denoting $\frac{\partial }{\partial z^{a}}$ as $\partial _{a}$, Eq. (\ref
{Veff-def}) yields
\begin{equation}
\partial _{a}V_{eff}=\frac{2V^{2}\partial _{a}V_{BH}-\left( \sqrt{1-4V_{BH}V}%
+2V_{BH}V-1\right) \partial _{a}V}{2V^{2}\sqrt{1-4V_{BH}V}},  \label{trtr}
\end{equation}
which further restricts the region of definition of $V_{eff}$ and $\partial
_{a}V_{eff}$ to $V_{BH}V<\frac{1}{4}$. In such a region, $V_{eff}$ also
enjoys the following power series expansion:
\begin{equation}
V_{eff}=\frac{1}{2V}+\frac{1}{2}\sum_{n=0}^{\infty }\binom{2n}{n}\frac{%
\left( V_{BH}\right) ^{n}V^{n-1}}{2n-1}=V_{BH}+\left( V_{BH}\right)
^{2}V+2\left( V_{BH}\right) ^{3}V^{2}+5\left( V_{BH}\right) ^{4}V^{3}+%
\mathcal{O}\left( \left( V_{BH}\right) ^{5}V^{4}\right) .
\end{equation}
\medskip

Let us finally point out that the extension of the formalism based on $%
V_{BH} $ to the presence of a non-vanishing scalar potential $V$, obtained
above, generally hold for a not necessarily supersymmetric theory described
by the action (\ref{action}), and does not rely on supersymmetry, and thus
on $N$, at all. It is also clear that one can easily generalize the
procedure exploited above for $d>4$ (for the case $d=5$, see also \cite
{ANYY1}).\medskip

However, when supersymmetry enters the game, one might ask about the
supersymmetry-preserving features of the critical points of $V_{eff}$. This
makes sense whenever the near-horizon geometry of the considered class of BH
backgrounds preserves some amount of supersymmetry, \textit{i.e.} when it
allows the supersymmetric variation of the (field-strength of the) gravitino
to vanish. In such a case, by studying the conditions of vanishing of the
supersymmetric variation of the gauginos, one can establish the possible
supersymmetry-preserving features of the scalar configuration corresponding
to the considered critical point of $V_{eff}$.

In the procedure performed above, we considered a dyonic, static,
spherically symmetric extremal BH background (\ref{met}), whose near-horizon
limit gives an $AdS_{2}\times S_{2}$ geometry. Such factor spaces
respectively have radii\ $r_{A}$ and $r_{H}$, which, as yielded by Eqs. (\ref
{CERNN2}) and (\ref{CERNN3}), in the presence of a non-vanishing gauging and
for $V_{BH}\neq 0$ do \textit{not} coincide. Thus, such a near-horizon
geometry is \textit{not} the Bertotti-Robinson (BR) one \cite{BR},
corresponding to $AdS_{2}\times S_{2}$ with $r_{A}=r_{H}$. Whereas BR
geometry is flat, conformally flat and it is a maximally supersymmetric $N=2$%
, $d=4$ background (see \textit{e.g.} \cite{Kallosh-BR} and \cite
{Kallosh-Peet-BR}, and Refs. therein), the near horizon limit (\ref
{near-horizon}) of the metric (\ref{met}) is neither flat nor conformally
flat; indeed its scalar curvature and non-vanishing components of Weyl
tensor can respectively computed to be
\begin{eqnarray}
&&R=2\left( \frac{1}{r_{A}^{2}}-\frac{1}{r_{H}^{2}}\right) ; \\
&&  \notag \\
&&\left\{
\begin{array}{l}
C_{trtr}=-\frac{1}{3}\left( \frac{1}{r_{A}^{2}}-\frac{1}{r_{H}^{2}}\right) ;
\\
\\
C_{t\theta t\theta }=-\frac{r^{2}}{6r_{A}^{4}}\left(
r_{H}^{2}-r_{A}^{2}\right) ; \\
\\
C_{t\varphi t\varphi }=-\frac{r^{2}}{6r_{A}^{4}}\left(
r_{H}^{2}-r_{A}^{2}\right) sin^{2}\theta ; \\
\\
C_{r\theta r\theta }=\frac{r_{A}^{2}-r_{H}^{2}}{6r^{2}}; \\
\\
C_{r\varphi r\varphi }=\frac{r_{A}^{2}-r_{H}^{2}}{6r^{2}}sin^{2}\theta ; \\
\\
C_{\theta \varphi \theta \varphi }=-\frac{r_{H}^{2}}{3r_{A}^{2}}\left(
r_{H}^{2}-r_{A}^{2}\right) sin^{2}\theta .
\end{array}
\right.
\end{eqnarray}

As shown in \cite{CKZ,Sabra1,Sabra2} (see also \cite{Fluxbranes}), in the
presence of a non-vanishing gauging, the vanishing of the $N=2$, $d=4$
supersymmetry variation of the gravitino in a static, spherically symmetric
extremal BH background holds only in the case of \textit{naked singularities}%
. Thus, one can conclude that the near horizon limit (\ref{near-horizon}) of
the metric (\ref{met}) (with $r_{A}\neq r_{H}$) does \textit{not} preserve
any supersymmetry at all. Consequently, it does not make sense to address
the issue of the supersymmetry-preserving features of the critical points of
$V_{eff}$ in such a framework. Necessarily, all scalar configurations
determining critical points of $V_{eff}$ will be \textit{non-supersymmetric}.

\section{\label{General-stu}Application to $stu$ Model with Fayet-Iliopoulos
Terms}

We are now going to apply the formalism and the general results obtained
above to a particular model of $N=2$, $d=4$ gauged supergravity, in which
the scalar potential $V(z,\bar{z})$ is the Fayet-Iliopoulos (FI) one (see
\textit{e.g.} \cite{N=2-big} and Refs. therein)\textbf{\ }
\begin{equation}
V_{FI}(z,\bar{z})\equiv (U^{\Lambda \Sigma }-3\bar{L}^{\Lambda }L^{\Sigma
})\xi _{\Lambda }^{x}\xi _{\Sigma }^{x},  \label{potscal}
\end{equation}
where
\begin{equation}
U^{\Lambda \Sigma }\equiv G^{a\overline{b}}\left( D_{a}L^{\Lambda }\right)
\overline{D}_{\overline{b}}\overline{L}^{\Lambda }=-\frac{1}{2}(\mu
)^{-1|\,\Lambda \Sigma }-\bar{L}^{\Lambda }L^{\Sigma },
\end{equation}
and $\xi _{\Lambda }^{x}$ are some real constants. This corresponds to the
gauging of a $U\left( 1\right) $, contained in the $\mathcal{R}$-symmetry $%
SU\left( 2\right) _{\mathcal{R}}$ and given by a particular moduli-dependent
alignment of the $n_{V}+1$ Abelian vector multiplets. Such a gauging does
not modify the derivatives of the scalar fields; moreover, in the considered
framework the hypermultiplets decouple, and thus we will not be dealing with
them. In this sense, we consider a gauging which is ``complementary'' to the
one recently studied in \cite{Vaula}.

By using some properties of the special K\"{a}hler geometry endowing the
vector multiplets' scalar manifold (see \textit{e.g.} \cite
{N=2-big,CDF-review} and Refs. therein), one obtains
\begin{equation}
D_{a}V_{FI}=\partial _{a}V_{FI}=\left[ iC_{abc}G^{b\overline{b}}G^{c%
\overline{c}}\left( \overline{D}_{\overline{b}}\overline{L}^{\Lambda
}\right) \overline{D}_{\overline{c}}\overline{L}^{\Lambda }-2\overline{L}%
^{\Lambda }D_{a}L^{\Sigma }\right] \xi _{\Lambda }^{x}\xi _{\Sigma }^{x}.
\end{equation}

We will consider the FI potential in the so-called $stu$ model,
based on the
factorized homogeneous symmetric special K\"{a}hler manifold ($dim_{\mathbb{C%
}}=n_{V}=3$) $\left( \frac{SU(1,1)}{U(1)}\right) ^{3}$ \cite
{Duff-stu,BKRSW,K3}, in three different frames of symplectic
coordinates. Remarkably, we will find that in the presence of
gauging (in our case, the
\textit{FI gauging} of the $U\left( 1\right) \subset SU\left( 2\right) _{\mathcal{R}}$%
) the symplectic invariance is lost, and one can obtain different
behavior of $V_{eff}$ (and thus different physics) by changing the
symplectic frame. This confirms the observation made long ago in the
framework of $N=4$, $d=4$ gauged supergravity in
\cite{Freedman-Schwarz}, and it is ultimately due to the fact that
in general, differently from $V_{BH}$, $V$ is not
symplectic-invariant.

\subsection{\label{stu}$\left( SO(1,1)\right) ^{2}$ Symplectic Basis}

The first basis we consider is the one manifesting the full $d=5$
isometry, \textit{i.e.} $\left( SO(1,1)\right) ^{2}$, which is a
particular
case, $n=2$, of the $SO\left( 1,1\right) \otimes SO\left( 1,n-1\right) $ ($%
n_{V}=n+1$) covariant basis \cite{Magnific-7}. In such a frame, the
holomorphic prepotential is simply $F=stu$ (with K\"{a}hler gauge
fixed such that $X^{0}\equiv 1$). The notation used for the
symplectic coordinates is as follows ($a=1,2,3$):
\begin{equation}
z^{a}=\{s\equiv x_{1}-ix_{2},\,t\equiv y_{1}-iy_{2},\,u\equiv z_{1}-iz_{2}\}.
\label{rainy2}
\end{equation}
This symplectic frame is obtained by the $N=2$ truncation of the
$USp\left( 8\right) $ gauged $N=8$ supergravity \cite{ADFL-gauging}.

Here and further below we will perform computations in the so-called \textit{%
magnetic} BH charge configuration\footnote{%
In Subsects. \ref{D2-D6} and \ref{D0-D6-2} we will treat also the case of
the so-called $D0-D6$ BH charge configuration.}, in which the non-vanishing
BH charges are $q_{0},\,p^{1},\,p^{2}$ and $\,p^{3}$. This does not imply
any loss of generality, due to the homogeneity of the manifold $\left( \frac{%
SU(1,1)}{U(1)}\right) ^{3}$ (as recently shown in \cite{GLS1}).

Within such a notation and BH charge configuration, the relevant geometrical
quantities look as follows (see \textit{e.g.} \cite{Ceresole})
\begin{eqnarray}
K &=&-\log \left( {8}x{_{2}\,}y{_{2}\,}z{_{2}}\right) ,\quad G_{a\overline{b}%
}=\frac{1}{4}diag[x_{2}^{-2},\,y_{2}^{-2},\,z_{2}^{-2}];  \label{CERNN1} \\
&&  \notag \\
Im\mathcal{N}_{\Lambda \Sigma } &=&-x_{2}\,y_{2}\,z_{2}\left(
\begin{array}{cccc}
1+\frac{x_{1}^{2}}{x_{2}^{2}}+\frac{y_{1}^{2}}{y_{2}^{2}}+\frac{z_{1}^{2}}{%
z_{2}^{2}} & -\frac{x_{1}}{x_{2}^{2}} & -\frac{y_{1}}{y_{2}^{2}} & -\frac{%
z_{1}^{2}}{z_{2}^{2}} \\
-\frac{x_{1}}{x_{2}^{2}} & \frac{1}{x_{2}^{2}} & 0 & 0 \\
-\frac{y_{1}}{y_{2}^{2}} & 0 & \frac{1}{y_{2}^{2}} & 0 \\
-\frac{z_{1}}{z_{2}^{2}} & 0 & 0 & \frac{1}{z_{2}^{2}}
\end{array}
\right) ;  \label{ImN-stu} \\
&&  \notag \\
Re\mathcal{N}_{\Lambda \Sigma } &=&\left(
\begin{array}{cccc}
x_{1}y_{1}z_{1} & -y_{1}z_{1} & -x_{1}z_{1} & -x_{1}y_{1} \\
-y_{1}z_{1} & 0 & z_{1} & y_{1} \\
-x_{1}z_{1} & z_{1} & 0 & x_{1} \\
-x_{1}y_{1} & y_{1} & x_{1} & 0
\end{array}
\right) ,  \label{ReN-stu}
\end{eqnarray}
and $V_{BH}$ reads
\begin{eqnarray}
V_{BH} &=&\frac{1}{2x_{2}y_{2}z_{2}}\left[
\begin{array}{l}
(q_{0})^{2}-2q_{0}(p^{1}y_{1}z_{1}+p^{2}x_{1}z_{1}+p^{3}x_{1}y_{1})+ \\
\\
+(p^{1})^{2}(y_{1}^{2}+y_{2}^{2})(z_{1}^{2}+z_{2}^{2})+(p^{2})^{2}(x_{1}^{2}+x_{2}^{2})(z_{1}^{2}+z_{2}^{2})+(p^{3})^{2}(x_{1}^{2}+x_{2}^{2})(y_{1}^{2}+y_{2}^{2})+
\\
\\
+2p^{1}p^{2}x_{1}y_{1}(z_{1}^{2}+z_{2}^{2})+2p^{1}p^{3}x_{1}z_{1}(y_{1}^{2}+y_{2}^{2})+2p^{2}p^{3}y_{1}z_{1}(x_{1}^{2}+x_{2}^{2})
\end{array}
\right] .  \notag \\
&&  \label{vbhSTU}
\end{eqnarray}

By using Eq. (\ref{potscal}), $V_{FI}$ is easily computed to be
\begin{equation}
V_{FI,\left( SO(1,1)\right) ^{2}}=-\frac{\xi _{2}^{x}\xi _{3}^{x}}{x_{2}}-%
\frac{\xi _{1}^{x}\xi _{3}^{x}}{y_{2}}-\frac{\xi _{1}^{x}\xi _{2}^{x}}{z_{2}}%
,  \label{scalpotstu}
\end{equation}
and $V_{eff,\left( SO(1,1)\right) ^{2}}$ can be calculated by recalling its
definition (\ref{Veff-def}).

It is worth pointing out that $V_{BH}$ is symplectic-invariant, and
thus Eq. (\ref{vbhSTU}) holds for the $stu$ model in the magnetic BH
charge configuration and \textit{in any symplectic frame}. On the
other hand, $V$
(whose particular from considered here is $V_{FI}$) and consequently $%
V_{eff} $ are \textit{not} symplectic-invariant. Thus, Eq. (\ref{scalpotstu}%
) and the corresponding expression of $V_{eff,\left( SO(1,1)\right)
^{2}}$ computable by using Eq. (\ref{Veff-def}) hold for the $stu$
model in the magnetic BH charge configuration \textit{only in the
}$\left( SO(1,1)\right) ^{2}$\textit{\ covariant basis}.

As one can see from Eq. (\ref{scalpotstu}), $V_{FI,\left( SO(1,1)\right)
^{2}}$ does not depend on the real parts of moduli. This implies that the
criticality conditions of $V_{eff,\left( SO(1,1)\right) ^{2}}$ with respect
to the real parts of moduli coincide with the analogous ones for $V_{BH}$:
\begin{equation}
\begin{array}{l}
\frac{\partial V_{eff,\left( SO(1,1)\right) ^{2}}}{\partial x^{1}}%
=0\Leftrightarrow \frac{\partial V_{BH}}{\partial x^{1}}=0; \\
\\
\frac{\partial V_{eff,\left( SO(1,1)\right) ^{2}}}{\partial y^{1}}%
=0\Leftrightarrow \frac{\partial V_{BH}}{\partial y^{1}}=0; \\
\\
\frac{\partial V_{eff,\left( SO(1,1)\right) ^{2}}}{\partial z^{1}}%
=0\Leftrightarrow \frac{\partial V_{BH}}{\partial z^{1}}=0.
\end{array}
\label{crit-crit}
\end{equation}
It is well known that criticality conditions on the right-hand side of Eq. (%
\ref{crit-crit}) imply in the BH magnetic charge configuration the \textit{%
``vanishing axions''} conditions $x_{1}=y_{1}=z_{1}=0$ (see \textit{e.g.}
\cite{Ceresole}), \textit{i.e.} they yield the purely imaginary nature of
the critical moduli. Thus, by using such \textit{on-shell} conditions for
the real part of the moduli, $V_{BH}$ can be rewritten in the following
partially \textit{on-shell}, \textit{reduced} form:
\begin{equation}
V_{BH,red}=\frac{1}{2x_{2}y_{2}z_{2}}\left[
(q_{0})^{2}+(p^{1})^{2}y_{2}^{2}z_{2}^{2}+(p^{2})^{2}x_{2}^{2}z_{2}^{2}+(p^{3})^{2}x_{2}^{2}y_{2}^{2}%
\right] .  \label{VBH-red}
\end{equation}
The symplectic invariance of the reduced forms of $V_{BH}$\ depends on the
symplectic invariance of the conditions implemented in order to go \textit{%
on-shell} for the variables dropped out. Due to Eqs. (\ref{crit-crit}), $%
V_{BH,red}$\ given by Eq. (\ref{VBH-red}) holds for the $stu$\ model
in the magnetic BH charge configuration and \textit{in any
symplectic frame}. By
using Eqs. (\ref{scalpotstu}) and (\ref{VBH-red}), one can also compute $%
V_{eff,\left( SO(1,1)\right) ^{2},red}$, whose expression clearly
holds only in the considered $\left( SO(1,1)\right) ^{2}$ covariant
basis.

\subsection{\label{Heterotic}$SO\left( 2,2\right) $ Symplectic Basis}

We now consider a symplectic frame exhibiting the maximum
non-compact symmetry $SO\left( 2,2\right) \sim \left( SU\left(
1,1\right) \right) ^{2}$; it is a particular case, $n=2$, of the
so-called Calabi-Visentini \cite {Calabi-Vis} $SO\left( 2,n\right) $
($n_{V}=n+1$) symplectic frame, considered in \cite{CDFVP}, having
heterotic stringy origin. The special K\"{a}hler geometry of $N=2$,
$d=4$ supergravity in such a symplectic frame is completely
specified by the following holomorphic symplectic section $\Omega $:
\begin{equation}
\Omega (u,s)=\left(
\begin{array}{c}
X^{\Lambda } \\
\\
F_{\Lambda }
\end{array}
\right) \equiv \left(
\begin{array}{c}
X^{\Lambda }(u) \\
\\
s\eta _{\Lambda \Sigma }X^{\Sigma }(u)
\end{array}
\right) ,  \label{Omega}
\end{equation}
where $\eta _{\Lambda \Sigma }=diag(1,1,-1,-1)$ and $X^{\Lambda }(u)$
satisfies the condition $X^{\Lambda }(u)\eta _{\Lambda \Sigma }X^{\Sigma
}(u)=0$. The axion-dilaton field $s\equiv x_{1}-ix_{2}$ parameterizes the
coset $\frac{SU(1,1)}{U(1)}$, whereas the two independent complex
coordinates $u_{1},u_{2}$ parameterize the coset $\frac{SO(2,2)}{SO(2)\times
SO(2)}$.

Note that, as shown in \cite{CDFVP}, in this symplectic frame a
prepotential does not exist at all. However, it is still possible to
calculate all the relevant geometrical quantities, using the
standard formul\ae\ of special K\"{a}hler geometry (see
\textit{e.g.} \cite{CDF-review} and Refs. therein). The K\"{a}hler
potential, the vector kinetic matrix and its real and imaginary
parts (along with the inverse of the imaginary part) respectively
read
\begin{align}
K& =-\log \left( {2x_{2}}\right) -\log \left( {X^{\Lambda }\eta _{\Lambda
\Sigma }\bar{X}^{\Sigma }}\right) ; \\
\mathcal{N}_{\Lambda \Sigma }& =-2ix_{2}(\bar{\Phi}_{\Lambda }\Phi _{\Sigma
}+\Phi _{\Lambda }\bar{\Phi}_{\Sigma })+(x_{1}+ix_{2})\eta _{\Lambda \Sigma
}; \\
\nu _{\Lambda \Sigma }& =x_{1}\eta _{\Lambda \Sigma }; \\
\mu _{\Lambda \Sigma }& =x_{2}\left[ \eta _{\Lambda \Sigma }-2(\bar{\Phi}%
_{\Lambda }\Phi _{\Sigma }+\Phi _{\Lambda }\bar{\Phi}_{\Sigma })\right] ; \\
(\mu ^{-1})^{\Lambda \Sigma }& =\frac{1}{x_{2}^{2}}\eta ^{\Lambda \Gamma
}\eta ^{\Sigma \Delta }\mu _{\Gamma \Delta }=\frac{1}{x_{2}}\left[ \eta
^{\Lambda \Sigma }-2(\bar{\Phi}^{\Lambda }\Phi ^{\Sigma }+\Phi ^{\Lambda }%
\bar{\Phi}^{\Sigma })\right] ,
\end{align}
where $\Phi ^{\Lambda }\equiv X^{\Lambda }/\sqrt{X^{\Lambda }\bar{X}%
_{\Lambda }}$ ($\Phi ^{\Lambda }\Phi _{\Lambda }=0$, $\Phi ^{\Lambda }\bar{%
\Phi}_{\Lambda }=1$), and all the indexes here and below are raised and
lowered by using $\eta _{\Lambda \Sigma }$ and its inverse $\eta ^{\Lambda
\Sigma }\equiv \left( \eta ^{-1}\right) ^{\Lambda \Sigma }=\eta _{\Lambda
\Sigma }$. By using such expressions and recalling the definition (\ref
{potscal}), one can calculate $V_{FI,SO\left( 2,2\right) }$ to be
\begin{equation}
V_{FI,SO\left( 2,2\right) }=-\frac{1}{2x_{2}}\xi _{\Lambda }^{x}\eta
^{\Lambda \Sigma }\xi _{\Sigma }^{x}.  \label{dom1}
\end{equation}

The relation between the $SO\left( 2,2\right) $ symplectic frame
specified by Eq. (\ref{Omega}) and the $\left( SO(1,1)\right) ^{2}$
symplectic frame is given by the formula \cite{CDFVP}:
\begin{equation}
X_{SO\left( 2,2\right) }^{\Lambda }=\frac{1}{\sqrt{2}}\left(
\begin{array}{c}
1-t\,u \\
-(t+u) \\
-(1+t\,u) \\
t-u
\end{array}
\right) .  \label{X-het-stu}
\end{equation}
Correspondingly, the BH charges in both symplectic frames are
related as follows \cite{CDFVP,BKRSW}
\begin{equation}
\left(
\begin{array}{c}
p^{\Lambda } \\
q_{\Lambda }
\end{array}
\right) _{SO\left( 2,2\right) }=\frac{1}{\sqrt{2}}\left(
\begin{array}{c}
p^{0}-q_{1} \\
-p^{2}-p^{3} \\
-p^{0}-q_{1} \\
p^{2}-p^{3} \\
p^{1}+q_{0} \\
-q_{2}-q_{3} \\
p^{1}-q_{0} \\
q_{2}-q_{3}
\end{array}
\right) _{\left( SO(1,1)\right) ^{2}}.  \label{charge}
\end{equation}

\subsection{\label{Duff-Liu}$SO\left( 8\right) $-truncated Symplectic Basis}

This symplectic frame, used in \cite{Duff} and recently in
\cite{Morales}, can be obtained from the $N=2$ truncation of the
$SO\left( 8\right) $ gauged $N=8$ supergravity \cite{Cremmer-Julia,
dWN}.

Let us start from the Lagrangian density\footnote{%
Whereas in \cite{Duff} the space-time signature $\left( -,+,+,+\right) $ was
used, we use $\left( +,-,-,-\right) $ instead.} (see Eq. (2.11) of \cite
{Duff})
\begin{eqnarray}
\mathcal{L} &=&\sqrt{-g}\left\{
\begin{array}{l}
-\frac{1}{2}R+\frac{1}{4}\left[ (\partial _{\mu }\phi ^{(12)})^{2}+(\partial
_{\mu }\phi ^{(13)})^{2}+(\partial _{\mu }\phi ^{(14)})^{2}\right] + \\
\\
-\left( e^{-\lambda _{1}}(\mathcal{F}_{\mu \nu }^{1})^{2}+e^{-\lambda _{2}}(%
\mathcal{F}_{\mu \nu }^{2})^{2}+e^{-\lambda _{3}}(\mathcal{F}_{\mu \nu
}^{3})^{2}+e^{-\lambda _{4}}(\mathcal{F}_{\mu \nu }^{4})^{2}\right)
-V_{FI,DL}
\end{array}
\right\} ,  \notag \\
&&  \label{lagduff}
\end{eqnarray}
where the $\lambda $s are defined as
\begin{eqnarray}
&&\lambda _{1}\equiv -\phi ^{(12)}-\phi ^{(13)}-\phi ^{(14)};  \notag \\
&&\lambda _{2}\equiv -\phi ^{(12)}+\phi ^{(13)}+\phi ^{(14)};  \notag \\
&&\lambda _{3}\equiv \phi ^{(12)}-\phi ^{(13)}+\phi ^{(14)};  \notag \\
&&\lambda _{4}\equiv \phi ^{(12)}+\phi ^{(13)}-\phi ^{(14)},
\end{eqnarray}
and the FI potential read
\begin{equation}
V_{FI}\equiv -2g^{2}\left( \cosh \phi ^{(12)}+\cosh \phi ^{(13)}+\cosh \phi
^{(14)}\right) .  \label{scalpotduf}
\end{equation}
By switching to new variables
\begin{equation}
\phi ^{(12)}\equiv \log \left( {X_{1}\,X_{2}}\right) ,\quad \phi
^{(13)}\equiv \log \left( {X_{1}\,X_{3}}\right) ,\quad \phi ^{(14)}\equiv
\log \left( {X_{1}\,X_{4}}\right) ,
\end{equation}
where the $X_{I}$ ($I=1,2,3,4$) are real and $X_{1}\,X_{2}\,X_{3}\,X_{4}=1$,
it is easy to rewrite the Lagrangian (\ref{lagduff}) into the form \cite
{Morales} (the subscript $SO\left( 8\right) $ stands for \textit{``}$%
SO\left( 8\right) $-truncated'')
\begin{equation}
\mathcal{L}_{SO\left( 8\right) }=\sqrt{-g}\left[ -\frac{1}{2}R+\frac{1}{4}%
X_{I}^{-2}\partial _{\mu }X_{I}\partial ^{\mu }X_{I}-X_{I}^{2}\mathcal{F}%
_{\mu \nu }^{I}\mathcal{F}^{I|\,\mu \nu }-V_{FI,SO\left( 8\right) }\right] ,
\label{lagmor}
\end{equation}
where
\begin{equation}
V_{FI,SO\left( 8\right) }\equiv -g^{2}\sum_{I<J}X_{I}X_{J}.  \label{VFI-MS}
\end{equation}
It is worth stressing that in the symplectic frame exploited by the
models above only electrical charges are non-vanishing.

The model studied in \cite{Morales} and based on the Lagrangian (\ref{lagmor}%
) (which, as shown above, is nothing but a rewriting of the model
considered in \cite{Duff} and based on the Lagrangian
(\ref{lagduff})) can be related to the $stu$ model in the symplectic
frame treated in Subsect. \ref{stu} in the following way.

Firstly, one has to perform a symplectic transformation from the magnetic BH
charge configuration to the purely electric configuration (thence
identifying $q_{1}\equiv p^{1}$, $q_{2}\equiv p^{2}$ and $q_{3}\equiv p^{3}$%
):
\begin{equation}
\left(
\begin{array}{c}
0 \\
0 \\
0 \\
0 \\
q_{0} \\
q_{1} \\
q_{2} \\
q_{3}
\end{array}
\right) =\mathcal{S}\left(
\begin{array}{c}
0 \\
p^{1} \\
p^{2} \\
p^{3} \\
q_{0} \\
0 \\
0 \\
0
\end{array}
\right) ,
\end{equation}
where $\mathcal{S}$ is an $8\times 8$ matrix ($\Lambda =0,1,2,3$) defined as
\begin{equation}
\mathcal{S}\equiv \left(
\begin{array}{cc}
A_{\Sigma }^{\Lambda } & \;B^{\Lambda \Sigma } \\
C_{\Lambda \Sigma } & \;D_{\Lambda }^{\Sigma }
\end{array}
\right) ,
\end{equation}
whose only non-vanishing components are $A_{0}^{0}=D_{0}^{0}=1$ and $%
B^{11}=B^{22}=B^{33}=-C_{11}=-C_{22}=-C_{33}=-1$.

This yields a corresponding symplectic transformations of sections of the $%
\left( SO(1,1)\right) ^{2}$ symplectic frame treated in Subsect.
\ref{stu}. By fixing the K\"{a}hler gauge such that $X^{0}\equiv 1$,
it reads
\begin{equation}
\mathcal{S}\left(
\begin{array}{c}
X^{\Lambda } \\
\\
F_{\Lambda }
\end{array}
\right) _{\left( SO(1,1)\right) ^{2},X^{0}\equiv 1}\mathcal{=S}\left(
\begin{array}{c}
1 \\
s \\
t \\
u \\
-stu \\
tu \\
su \\
st
\end{array}
\right) =\left(
\begin{array}{c}
1 \\
-tu \\
-su \\
-st \\
-stu \\
s \\
t \\
u
\end{array}
\right) \equiv \left(
\begin{array}{c}
X^{\Lambda } \\
\\
F_{\Lambda }
\end{array}
\right) _{SO\left( 8\right) ,X^{0}\equiv 1}.
\end{equation}
One can also compute that
\begin{equation}
F_{SO\left( 8\right) ,X^{0}\equiv 1}=-2stu=-2F_{\left( SO(1,1)\right)
^{2},X^{0}\equiv 1},
\end{equation}
or, in terms of the relevant $X^{\Lambda }$s\footnote{%
Notice that $X_{\left( SO\left( 1,1\right) \right) ^{2}}^{0}=X_{SO\left(
8\right) }^{0}\equiv X^{0}$, because the symplectic transformation $\mathcal{%
S}$ does not act on $X^{0}$.}:
\begin{equation}
F_{SO\left( 8\right) ,X^{0}\equiv 1}=-2\left. \sqrt{-X^{0}X_{SO\left(
8\right) }^{1}X_{SO\left( 8\right) }^{2}X_{SO\left( 8\right) }^{3}}\right|
_{X^{0}\equiv 1}=-2\left. \frac{X_{\left( SO(1,1)\right) ^{2}}^{1}X_{\left(
SO(1,1)\right) ^{2}}^{2}X_{\left( SO(1,1)\right) ^{2}}^{3}}{X^{0}}\right|
_{X^{0}\equiv 1}.
\end{equation}

Finally, by recalling the symplectic-invariant \textit{``vanishing axions''}
conditions (\ref{crit-crit}) and the definitions (\ref{rainy2}), one can
introduce the following holomorphic sections:
\begin{equation}
\left(
\begin{array}{c}
\widehat{X}^{\Lambda } \\
\\
\widehat{F}_{\Lambda }
\end{array}
\right) _{SO\left( 8\right) ,X^{0}\equiv 1}\equiv \left(
\begin{array}{c}
X^{\Lambda } \\
\\
F_{\Lambda }
\end{array}
\right) _{SO\left( 8\right) ,X^{0}\equiv 1,x_{1}=y_{1}=z_{1}=0}=\left(
\begin{array}{c}
1 \\
y_{2}z_{2} \\
x_{2}z_{2} \\
x_{2}y_{2} \\
-ix_{2}y_{2}z_{2} \\
-ix_{2} \\
-iy_{2} \\
-iz_{2}
\end{array}
\right) ,  \label{DL-sections}
\end{equation}
where $x_{2}$, $y_{2}$ and $z_{2}$ have been defined in Subsect. \ref{stu}.

By using the general transformation rule for the matrix $\mathcal{N}$ (see
\textit{e.g.} \cite{Ceresole1}) and the explicit expressions (\ref{ImN-stu})
and (\ref{ReN-stu}) of its imaginary and real parts in the $stu$ model, one
can easily compute the kinetic vector matrix corresponding to sections (\ref
{DL-sections}), denoted by $\widehat{\mathcal{N}}$. Its real part vanishes,
while its imaginary reads as follows:
\begin{equation}
\hat{\mu}_{00}=-x_{2}y_{2}z_{2},\quad \hat{\mu}_{11}=-\frac{x_{2}}{y_{2}z_{2}%
},\quad \hat{\mu}_{22}=-\frac{y_{2}}{z_{2}x_{2}},\quad \hat{\mu}_{33}=-\frac{%
z_{2}}{x_{2}y_{2}}.  \label{DL-N}
\end{equation}
By inserting Eqs. (\ref{DL-sections}) and (\ref{DL-N}) into the definition (%
\ref{potscal}), one obtains the following expression of the FI potential:
\begin{equation}
V_{FI,SO\left( 8\right) }=-\xi _{0}^{x}(\frac{\xi _{1}^{x}}{x_{2}}+\frac{\xi
_{2}^{x}}{y_{2}}+\frac{\xi _{3}^{x}}{z_{2}})-\xi _{2}^{x}\xi
_{3}^{x}x_{2}-\xi _{1}^{x}\xi _{3}^{x}y_{2}-\xi _{1}^{x}\xi _{2}^{x}z_{2}.
\label{admor}
\end{equation}
As anticipated in the Introduction, $V_{FI,SO\left( 8\right) }$ is the only
gauge potential (among the considered ones) admitting critical points at a
finite distance from the origin. By comparing Eq. (\ref{admor}) with Eqs. (%
\ref{scalpotstu}) and (\ref{dom1}) above, it is easy to realize that this is
due to the simultaneous presence of \textit{``}$\frac{1}{x}$'' and \textit{``%
}$x$'' terms, thus allowing for a \textit{``}balance'' of opposite behaviors
of the potential.

Comparing the matrix $\hat{\mu}_{\Lambda \Sigma }$ given by Eq. (\ref{DL-N})
with the one considered in \cite{Morales}, namely with ($I=1,2,3,4$)
\begin{equation}
\mu _{IJ}=-X_{I}^{2}\,\delta _{IJ},\quad \nu _{IJ}=0,
\end{equation}
one can relate the $\left( SO(1,1)\right) ^{2}$ symplectic coordinates with
the ones used in \cite{Morales}:
\begin{equation}
X_{1}^{2}=\frac{x_{2}}{y_{2}z_{2}},\quad X_{2}^{2}=\frac{y_{2}}{x_{2}z_{2}}%
,\quad X_{3}^{2}=\frac{z_{2}}{y_{2}x_{2}},\quad X_{4}^{2}=x_{2}y_{2}z_{2}.
\end{equation}
It is worth pointing out that the FI potential (\ref{VFI-MS}) of \cite
{Morales} corresponds to the FI potential (\ref{admor}) with the particular
choice $\xi _{0}^{x}=\xi _{1}^{x}=\xi _{2}=\xi _{3}^{x}=\sqrt{2g^{2}}$.

By recalling that $V_{BH,red}$ is symplectic invariant and given by Eq. (\ref
{VBH-red}), and using Eqs. (\ref{Veff-def}) and (\ref{admor}), one can then
compute $V_{eff,red,SO\left( 8\right) }$.

\section{\label{BH-Entropies}Dependence of Black Hole Entropy on the Symplectic Frame in $stu$ Model}

We will now go \textit{on-shell} for the formalism based on the effective BH
potential $V_{eff}$. In other words, we will compute, for the considered
magnetic BH charge configuration, the critical points of $V_{eff}$ of the $%
stu$ model in the three symplectic frames introduced above, and then
compare the resulting BH entropies. In the $SO\left( 8\right)
$-truncated symplectic basis we will match perfectly the results of
\cite{Morales}, obtained by using the (intrinsically
\textit{on-shell}) Sen's\textit{\ entropy function formalism} (see
\cite{Sen-old1,Sen-old2}, and \cite{Sen-review} and Refs. therein).
Furthermore, as previously mentioned, the formal and physical
results will generally depend on the symplectic basis being
considered. We will also consider the case of $D0$-$D6$ BH charge
configuration.

\subsection{\label{Magnetic}Magnetic Configuration}

It is now convenient to introduce the following new coordinates:
\begin{equation}
x_{2}\equiv \sqrt{\left| \frac{q_{0}p^{1}}{p^{2}p^{3}}\right| }\,x,\quad
y_{2}\equiv \sqrt{\left| \frac{q_{0}p^{2}}{p^{1}p^{3}}\right| }\,y,\quad
z_{2}\equiv \sqrt{\left| \frac{q_{0}p^{3}}{p^{2}p^{1}}\right| }\,z.
\label{male-male}
\end{equation}
By doing so, Eqs. (\ref{VBH-red}), (\ref{scalpotstu}), (\ref{dom1})
and (\ref {admor}), holding in the magnetic BH charge configuration
and expressing the symplectic-invariant $V_{BH}$ and the not
symplectic-invariant $V_{FI}$ in the $\left( SO(1,1)\right) ^{2}$,
$SO\left( 2,2\right) $ covariant and $SO\left( 8\right) $-truncated
symplectic frames respectively, can be rewritten as follows:
\begin{equation}
\left\{
\begin{array}{l}
V_{BH,red}=\frac{\sqrt{|q_{0}p^{1}p^{2}p^{3}|}}{2}\,\widetilde{V}_{BH,red},
\\
\\
\,\widetilde{V}_{BH,red}\equiv \frac{1+x^{2}y^{2}+x^{2}z^{2}+y^{2}z^{2}}{xyz}%
;
\end{array}
\right.  \label{VBHSTU1}
\end{equation}
\begin{eqnarray}
&&\left\{
\begin{array}{l}
V_{FI,\left( SO(1,1)\right) ^{2}}=-\frac{1}{2\sqrt{|q_{0}p^{1}p^{2}p^{3}|}}\,%
\tilde{V}_{FI,\left( SO(1,1)\right) ^{2}}, \\
\\
\tilde{V}_{FI,\left( SO(1,1)\right) ^{2}}\equiv \frac{\tilde{\xi}_{2}\tilde{%
\xi}_{3}}{x}+\frac{\tilde{\xi}_{1}\tilde{\xi}_{3}}{y}+\frac{\tilde{\xi}_{1}%
\tilde{\xi}_{2}}{z}, \\
\\
\tilde{\xi}_{i}\equiv \sqrt{2}|p^{i}|\xi _{i}^{x},i=1,2,3;
\end{array}
\right.  \label{scalpotstu1} \\
&&  \notag \\
&&\left\{
\begin{array}{l}
V_{FI,SO\left( 2,2\right) }=-\frac{1}{2\sqrt{|q_{0}p^{1}p^{2}p^{3}|}}\,%
\tilde{V}_{SO\left( 2,2\right) }, \\
\\
\tilde{V}_{FI,SO\left( 2,2\right) }\equiv \frac{\tilde{\alpha}}{x}, \\
\\
\widetilde{\alpha }\equiv |p^{2}p^{3}|\xi _{\Lambda }^{x}\eta ^{\Lambda
\Sigma }\xi _{\Sigma }^{x};
\end{array}
\right.  \label{scalpothet1} \\
&&  \notag \\
&&\left\{
\begin{array}{l}
V_{FI,SO\left( 8\right) }=-\frac{1}{2\sqrt{|q_{0}p^{1}p^{2}p^{3}|}}\,%
\widetilde{V}_{FI,SO\left( 8\right) }, \\
\\
\widetilde{V}_{FI,SO\left( 8\right) }\equiv \tilde{\xi}_{0}(\frac{\tilde{\xi}%
_{1}}{x}+\frac{\tilde{\xi}_{2}}{y}+\frac{\tilde{\xi}_{3}}{z})+\tilde{\xi}_{2}%
\tilde{\xi}_{3}x+\tilde{\xi}_{1}\tilde{\xi}_{3}y+\tilde{\xi}_{1}\tilde{\xi}%
_{2}z, \\
\\
\tilde{\xi}_{0}\equiv \sqrt{\left| 2\frac{p^{1}p^{2}p^{3}}{q_{0}}\right| }%
\,\xi _{0}^{x},\tilde{\xi}_{1}\equiv \sqrt{\left| 2\frac{q_{0}p^{2}p^{3}}{%
p^{1}}\right| }\,\xi _{1}^{x},\tilde{\xi}_{2}\equiv \sqrt{\left| 2\frac{%
q_{0}p^{1}p^{3}}{p^{2}}\right| }\,\xi _{2}^{x},\tilde{\xi}_{3}\equiv \sqrt{%
\left| 2\frac{q_{0}p^{1}p^{2}}{p^{3}}\right| }\,\xi _{3}^{x}.
\end{array}
\right.  \label{scalpotmor1}
\end{eqnarray}

Let us also write here the expressions of the (symplectic-invariant)
superpotential and of its covariant derivatives (\textit{matter charges}):
\begin{eqnarray}
&&W=q_{0}\left\{ 1+sgn\left( q_{0}\right) \left[ sgn\left( p^{1}\right)
yz+sgn\left( p^{2}\right) xz+sgn\left( p^{3}\right) xy\right] \right\} ;
\label{trag1} \\
&&D_{x}W=-\frac{q_{0}}{x}\left[ 1+sgn\left( q_{0}p^{1}\right) yz\right] ; \\
&&D_{y}W=-\frac{q_{0}}{y}\left[ 1+sgn\left( q_{0}p^{2}\right) xz\right] ; \\
&&D_{z}W=-\frac{q_{0}}{z}\left[ 1+sgn\left( q_{0}p^{3}\right) xy\right] .
\label{trag2}
\end{eqnarray}

Now, by recalling the definition (\ref{Veff-def}), the reduced effective
potential can be written as follows:
\begin{equation}
\left\{
\begin{array}{l}
V_{eff,red}=\sqrt{|q_{0}p^{1}p^{2}p^{3}|}\,\tilde{V}_{eff,red}; \\
\\
\tilde{V}_{eff,red}\equiv \frac{\sqrt{1+\tilde{V}_{BH,red}\tilde{V}_{FI}}-1}{%
\tilde{V}_{FI}},
\end{array}
\right.
\end{equation}
where $\tilde{V}_{BH,red}$ is given by Eq. (\ref{VBHSTU1}), and $\tilde{V}%
_{FI}$, depending on the symplectic frame, is given by Eqs. (\ref
{scalpotstu1})-(\ref{scalpotmor1}).

\subsubsection{\label{lundi1}$SO\left( 2,2\right) $}

$\tilde{V}_{eff,red,SO\left( 2,2\right) }$ reads
\begin{equation}
\tilde{V}_{eff,red,SO\left( 2,2\right) }=x\frac{\sqrt{1+\frac{\tilde{\alpha}%
}{x^{2}yz}(1+x^{2}y^{2}+x^{2}z^{2}+y^{2}z^{2})}-1}{\tilde{\alpha}}.
\end{equation}
Attractor Eqs. for such a potential yield the following solutions:
\begin{equation}
\frac{\partial \tilde{V}_{eff,red,SO\left( 2,2\right) }}{\partial x}=\frac{%
\partial \tilde{V}_{eff,red,SO\left( 2,2\right) }}{\partial y}=\frac{%
\partial \tilde{V}_{eff,red,SO\left( 2,2\right) }}{\partial z}%
=0\Leftrightarrow \left\{
\begin{array}{l}
\tilde{\alpha}>-\frac{1}{2}:x=\frac{1}{\sqrt{1+2\tilde{\alpha}}},\quad
y=z=\pm 1; \\
\\
\tilde{\alpha}<\frac{1}{2}:x=-\frac{1}{\sqrt{1-2\tilde{\alpha}}},\quad
y=-z=\pm 1,
\end{array}
\right.  \label{male!-1}
\end{equation}
and the corresponding BH entropy reads
\begin{equation}
S_{BH,SO\left( 2,2\right) }=\left\{
\begin{array}{l}
\tilde{\alpha}>-\frac{1}{2}:2\pi \frac{\sqrt{|q_{0}p^{1}p^{2}p^{3}|}}{\sqrt{%
1+2\tilde{\alpha}}}; \\
\\
\tilde{\alpha}<\frac{1}{2}:2\pi \frac{\sqrt{|q_{0}p^{1}p^{2}p^{3}|}}{\sqrt{%
1-2\tilde{\alpha}}}.
\end{array}
\right.
\end{equation}
As it can be seen by recalling Eqs. (\ref{trag1})-(\ref{trag2}), for both
ranges of $\tilde{\alpha}$ the solutions have at least one non-vanishing
matter charge. The condition $Z=0$ requires $p^{2}p^{3}<0$ and $p^{1}q_{0}<0$%
, whereas $Z\neq 0$ requires $p^{2}p^{3}>0$ and $p^{1}q_{0}<0$.

The ungauged limit $\tilde{\alpha}\rightarrow 0$ yields $x=\pm 1$ and $%
S_{BH,SO\left( 2,2\right) }=2\pi \sqrt{\left| q_{0}p^{1}p^{2}p^{3}\right| }$%
, consistenty with the known results \cite{Shmakova,TT,TT2}. On the other
hand, by recalling Eq. (\ref{CERNN1}) the limit $\tilde{\alpha}\rightarrow
\infty $ in the branch $\tilde{\alpha}>-\frac{1}{2}$ and the limit $\tilde{%
\alpha}\rightarrow -\infty $ in the branch $\tilde{\alpha}<\frac{1}{2}$
yield a singular geometry.

\subsubsection{\label{lundi2}$\left( SO(1,1)\right) ^{2}$}

As it can be seen from Eqs. (\ref{scalpotstu1}) and
(\ref{scalpothet1}), in the case $\xi _{1}=0$ one obtains that
$V_{FI,\left( SO(1,1)\right) ^{2}}\varpropto V_{FI,SO\left(
2,2\right) }$, and everything, up to some redefinitions, goes as in
the $SO\left( 2,2\right) $ symplectic basis.

In the following treatment we will assume the simplifying assumption
(implying asymptotically $AdS$ BHs)
\begin{equation}
\tilde{\xi}_{1}=\tilde{\xi}_{2}=\tilde{\xi}_{3}=\sqrt{\frac{g^{2}}{3}},
\label{lunnn1}
\end{equation}
corresponding to the most symmetric implementation of the \textit{FI
gauging} in the considered symplectic frame. Under such an
assumption, one gets
\begin{equation}
\tilde{V}_{eff,red,\left( SO(1,1)\right) ^{2}}=\frac{\sqrt{x^{2}y^{2}z^{2}+%
\frac{g^{2}}{3}(xy+xz+yz)(1+x^{2}y^{2}+x^{2}z^{2}+y^{2}z^{2})}-xyz}{\frac{%
g^{2}}{3}(xy+xz+yz)}.
\end{equation}
Attractor Eqs. for such a potential yield the following solution:
\begin{gather}
\frac{\partial \tilde{V}_{eff,red,\left( SO(1,1)\right) ^{2}}}{\partial x}=%
\frac{\partial \tilde{V}_{eff,red,\left( SO(1,1)\right) ^{2}}}{\partial y}=%
\frac{\partial \tilde{V}_{eff,red,\left( SO(1,1)\right) ^{2}}}{\partial z}=0;
\notag \\
\Updownarrow  \notag \\
x=y=z=\left( \frac{3+6g^{2}+\sqrt{9+24g^{2}}}{6+18g^{2}}\right) ^{1/4}.
\label{tragedia1}
\end{gather}

The corresponding BH entropy reads
\begin{equation}
S_{BH,\left( SO(1,1)\right) ^{2}}=\pi \sqrt{|q_{0}p^{1}p^{2}p^{3}|}\frac{%
\left[ -\sqrt{2}+\sqrt{5+12g^{2}+\sqrt{9+24g^{2}}}\right] }{2g^{2}}\left(
\frac{6+12g^{2}+2\sqrt{9+24g^{2}}}{3+9g^{2}}\right) ^{1/4}.  \label{traggg}
\end{equation}
In the ungauged limit $g\rightarrow 0$ one gets $x=y=z=1$ and the well-known
expression of the BH entropy is recovered \cite{Shmakova,TT,TT2}:
\begin{equation}
lim_{g\rightarrow 0}S_{BH,\left( SO(1,1)\right) ^{2}}=2\pi \sqrt{%
|q_{0}p^{1}p^{2}p^{3}|}.
\end{equation}
For $g\rightarrow \infty $ $S_{BH,\left( SO(1,1)\right) ^{2}}$ vanishes and $%
x=y=z=3^{-1/4}$. As it can be seen by recalling Eqs. (\ref{trag1})-(\ref
{trag2}), the solution (\ref{tragedia1}) has all $Z$, $D_{x}Z$, $D_{y}Z$ and
$D_{z}Z$ non-vanishing.

\subsubsection{\label{lundi3}$SO\left( 8\right) $-truncated}

As it can be seen from Eqs. (\ref{scalpotmor1}) and (\ref{scalpothet1}), in
the case $\tilde{\xi}_{2}=\tilde{\xi}_{3}=0$ one obtains that $%
V_{FI,SO\left( 8\right) }\varpropto V_{FI,SO\left( 2,2\right) }$,
and everything, up to some redefinitions, goes as in the $SO\left(
2,2\right) $ symplectic basis.

In the following treatment we will assume the simplifying assumption
\begin{equation}
\tilde{\xi}_{0}=\tilde{\xi}_{1}=\tilde{\xi}_{2}=\tilde{\xi}_{3},
\label{monday-1}
\end{equation}
corresponding to the most symmetric implementation of the \textit{FI
gauging} in the
considered symplectic frame. Under such an assumption, one gets\footnote{%
Here and below $\tilde{\alpha}_{SO\left( 8\right) }\geqslant 0$, implying
asymptotically $AdS$ (flat in the ungauged limit $\tilde{\alpha}_{SO\left(
8\right) }\rightarrow 0$) BHs.}
\begin{equation}
\tilde{V}_{eff,red,SO\left( 8\right) }=\frac{\sqrt{1+\frac{\tilde{\alpha}%
_{SO\left( 8\right) }}{x^{3}}(1+3x^{4})(\frac{1}{x}+x)}-1}{\tilde{\alpha}%
_{SO\left( 8\right) }(1/x+x)},  \label{chesucc1}
\end{equation}
where (without loss of generality, under the assumption (\ref{monday-1})) we
put $x=y=z$, and $\tilde{\alpha}_{SO\left( 8\right) }\equiv 3\tilde{\xi}%
_{0}^{2}$.

The attractor Eq. for such a potential yields the following $\tilde{\alpha}%
_{SO\left( 8\right) }$-independent solution:
\begin{equation}
\frac{\partial \tilde{V}_{eff,red,SO\left( 8\right) }}{\partial x}%
=0\Leftrightarrow x=1,  \label{male!2}
\end{equation}
and the corresponding BH entropy reads
\begin{equation}
S_{BH,SO\left( 8\right) }=\pi \sqrt{|q_{0}p^{1}p^{2}p^{3}|}\frac{\sqrt{1+8%
\tilde{\alpha}_{SO\left( 8\right) }}-1}{2\tilde{\alpha}_{SO\left( 8\right) }}%
.
\end{equation}
Such an expression coincides with the result obtained in \cite{Morales}, up
to a redefinition of the constant $\tilde{\alpha}_{SO\left( 8\right) }$ by
considering $q_{0}=p^{1}=p^{2}=p^{3}$. Consistently, in the ungauged limit $%
\tilde{\alpha}_{SO\left( 8\right) }\rightarrow 0$ the well known expression
of the BH entropy is recovered \cite{Shmakova,TT,TT2}:
\begin{equation}
lim_{g\rightarrow 0}S_{BH,SO\left( 8\right) }=2\pi \sqrt{%
|q_{0}p^{1}p^{2}p^{3}|},
\end{equation}
whereas $S_{BH,SO\left( 8\right) }$ vanishes in the limit $\tilde{\alpha}%
_{SO\left( 8\right) }\rightarrow \infty $.

As it can be seen by recalling Eqs. (\ref{trag1})-(\ref{trag2}), depending
on the sign of charges can make $Z$ or $D_{x}Z$, $D_{y}Z$ and $D_{z}Z$
vanish, or also obtain the non-vanishing of all $Z$, $D_{x}Z$, $D_{y}Z$ and $%
D_{z}Z$.

\subsection{\label{D2-D6}$D0$-$D6$ Configuration}

Let us now consider the so-called $D0$-$D6$ BH charge configuration,
in which (in the $\left( SO(1,1)\right) ^{2}$ covariant symplectic
basis) all charges vanish but $p^{0}$ and $\,q_{0}$. For such a
configuration the symplectic-invariant $V_{BH}$, defined by Eq.
(\ref{VBH-def}), reads
\begin{equation}
V_{BH,D0-D6}=\frac{1}{2x_{2}y_{2}z_{2}}\left[
(q_{0})^{2}+2q_{0}p^{0}x_{1}y_{1}z_{1}+p_{0}^{2}(x_{1}^{2}+x_{2}^{2})(y_{1}^{2}+y_{2}^{2})(z_{1}^{2}+z_{2}^{2})%
\right] .
\end{equation}
By recalling the expressions of the FI potentials given by Eqs. (\ref
{scalpotstu}), (\ref{dom1}) and (\ref{admor}), it is easy to realize that
the criticality conditions of $V_{eff,D0-D6}$ with respect to the real parts
of moduli coincide with the analogous ones for $V_{BH,D0-D6}$:
\begin{equation}
\begin{array}{c}
\frac{\partial V_{eff,D0-D6}}{\partial x^{1}}=0\Leftrightarrow \frac{%
\partial V_{BH,D0-D6}}{\partial x^{1}}=0; \\
\\
\frac{\partial V_{eff,D0-D6}}{\partial y^{1}}=0\Leftrightarrow \frac{%
\partial V_{BH,D0-D6}}{\partial y^{1}}=0; \\
\\
\frac{\partial V_{eff,D0-D6}}{\partial z^{1}}=0\Leftrightarrow \frac{%
\partial V_{BH,D0-D6}}{\partial z^{1}}=0.
\end{array}
\label{crit-crit-bad}
\end{equation}
It is well known that criticality conditions on the right-hand side of Eq. (%
\ref{crit-crit-bad}) imply in the $D0-D6$ BH charge configuration the
conditions $x_{1}=y_{1}=z_{1}=0$ (see \textit{e.g.} \cite{Ceresole}),
\textit{i.e.} they yield the purely imaginary nature of the critical moduli.
Thus, by using such \textit{on-shell} conditions for the real part of the
moduli, $V_{BH,D0-D6}$ can be rewritten in the following partially \textit{%
on-shell}, \textit{reduced} form:
\begin{equation}
V_{BH,D0-D6,red}=\frac{1}{2x_{2}y_{2}z_{2}}\left[
(q_{0})^{2}+(p^{0})^{2}x_{2}^{2}y_{2}^{2}z_{2}^{2}\right] .  \label{lun1}
\end{equation}
As noticed above, the symplectic invariance of the reduced forms of $V_{BH}$%
\ depends on the symplectic invariance of the conditions implemented in
order to go \textit{on-shell} for the variables dropped out. Due to Eqs. (%
\ref{crit-crit-bad}), $V_{BH,D0-D6,red}$\ given by Eq. (\ref{lun1})
holds for the $stu$\ model in the $D0-D6$ BH charge configuration
and in any symplectic frame.

It is now convenient to introduce the following new coordinates:
\begin{equation}
x_{2}\equiv \sqrt[3]{\left| \frac{q_{0}}{p^{0}}\right| }\,x,\quad
y_{2}\equiv \sqrt[3]{\left| \frac{q_{0}}{p^{0}}\right| }\,y,\quad
z_{2}\equiv \sqrt[3]{\left| \frac{q_{0}}{p^{0}}\right| }\,z.
\end{equation}
By doing so, Eqs. (\ref{lun1}), (\ref{scalpotstu}), (\ref{dom1}) and (\ref
{admor}) can respectively be rewritten as follows (we drop the subscript $%
\mathit{D0-D6}$):
\begin{equation}
\left\{
\begin{array}{l}
V_{BH,red}=\frac{|q_{0}p^{0}|}{2}\tilde{V}_{BH,red}, \\
\\
\tilde{V}_{BH,red}\equiv \frac{1+x^{2}y^{2}z^{2}}{xyz};
\end{array}
\right.  \label{vbhp0q01}
\end{equation}
and
\begin{eqnarray}
&&\left\{
\begin{array}{l}
V_{FI,\left( SO(1,1)\right) ^{2}}=-\frac{1}{2|q_{0}p^{0}|}\,\tilde{V}%
_{FI,\left( SO(1,1)\right) ^{2}}, \\
\\
\tilde{V}_{FI,\left( SO(1,1)\right) ^{2}}\equiv \frac{\tilde{\xi}_{2}\tilde{%
\xi}_{3}}{x}+\frac{\tilde{\xi}_{1}\tilde{\xi}_{3}}{y}+\frac{\tilde{\xi}_{1}%
\tilde{\xi}_{2}}{z}, \\
\\
\tilde{\xi}_{i}\equiv \sqrt{2}\sqrt[3]{(p^{0})^{2}|q_{0}|}\,\xi
_{i}^{x},i=1,2,3;
\end{array}
\right.  \label{scalpotstu2} \\
&&  \notag \\
&&\left\{
\begin{array}{l}
V_{FI,SO\left( 2,2\right) }=-\frac{1}{2|q_{0}p^{0}|}\,\tilde{V}_{FI,SO\left(
2,2\right) }, \\
\\
\tilde{V}_{FI,SO\left( 2,2\right) }\equiv \frac{\tilde{\alpha}}{x}, \\
\\
\widetilde{\alpha }\equiv \sqrt[3]{(p^{0})^{4}q_{0}^{2}}\,\xi _{\Lambda
}^{x}\eta ^{\Lambda \Sigma }\xi _{\Sigma }^{x};
\end{array}
\right.  \label{scalpothet2} \\
&&  \notag \\
&&\left\{
\begin{array}{l}
V_{FI,SO\left( 8\right) }=-\frac{1}{2|q_{0}p^{0}|}\,\tilde{V}_{FI,SO\left(
8\right) }, \\
\\
\tilde{V}_{FI,SO\left( 8\right) }\equiv \tilde{\xi}_{0}(\frac{\tilde{\xi}_{1}%
}{x}+\frac{\tilde{\xi}_{2}}{y}+\frac{\tilde{\xi}_{3}}{z})+\tilde{\xi}_{2}%
\tilde{\xi}_{3}x+\tilde{\xi}_{1}\tilde{\xi}_{3}y+\tilde{\xi}_{1}\tilde{\xi}%
_{2}z, \\
\\
\tilde{\xi}_{0}\equiv \sqrt{2}|p^{0}|\,\xi _{0}^{x},\tilde{\xi}_{i}\equiv
\sqrt{2}\sqrt[3]{(q_{0})^{2}|p^{0}|}\,\xi _{i}^{x},i=1,2,3.
\end{array}
\right.  \label{scalpotmor2}
\end{eqnarray}
Let us also write here the expressions of the (symplectic-invariant)
superpotential and of its covariant derivatives (\textit{matter charges}):
\begin{eqnarray}
&&W=q_{0}\left[ 1+sgn\left( q_{0}p^{0}\right) xyz\right] ;  \label{lun2} \\
&&D_{x}W=-\frac{q_{0}}{x}; \\
&&D_{y}W=-\frac{q_{0}}{y}; \\
&&D_{z}W=-\frac{q_{0}}{z}.  \label{lun3}
\end{eqnarray}

Now, by recalling the definition (\ref{Veff-def}), the effective potential
can be written as follows:
\begin{equation}
\left\{
\begin{array}{l}
V_{eff,red}=|q_{0}p^{0}|\,\tilde{V}_{eff,red}, \\
\\
\tilde{V}_{eff,red}\equiv \frac{\sqrt{1+\tilde{V}_{BH,red}\tilde{V}_{FI}}-1}{%
\tilde{V}_{FI}},
\end{array}
\right.
\end{equation}
where $\tilde{V}_{BH,red}$ is given by Eq. (\ref{vbhp0q01}), and $\tilde{V}%
_{FI}$, depending on the symplectic basis, is given by Eqs. (\ref
{scalpotstu2})-(\ref{scalpotmor2}).

\subsubsection{\label{lundi4}$SO\left( 2,2\right) $}

$\tilde{V}_{eff,red,SO\left( 2,2\right) }$ reads
\begin{equation}
\tilde{V}_{eff,red,SO\left( 2,2\right) }=x\frac{\sqrt{1+\frac{\tilde{\alpha}%
}{x^{2}yz}(1+x^{2}y^{2}z^{2})}-1}{\tilde{\alpha}}.
\end{equation}
The corresponding attractor Eqs. for the variables $y$ and $z$ yield:
\begin{equation}
\frac{\partial \tilde{V}_{eff,red,SO\left( 2,2\right) }}{\partial y}=\frac{%
\partial \tilde{V}_{eff,red,SO\left( 2,2\right) }}{\partial z}%
=0\Leftrightarrow xyz=1.  \label{lunn1}
\end{equation}
On the other hand, when inserting the attractor Eq. for the variable $x$
into the condition (\ref{lunn1}) one obtains
\begin{equation}
\widetilde{\alpha }yz=0.  \label{lunn2}
\end{equation}
It is then easy to realize that the attractor Eqs. have solutions only in
the ungauged case $\widetilde{\alpha }=0$. In such a case, the corresponding
solution is non-BPS $Z\neq 0$, it is stable (up to a non-BPS $Z\neq 0$
moduli space with $dim_{\mathbb{R}}=2$ \cite{BFGM1,TT2,fm07}), and the
corresponding BH entropy is given by the well known result \cite
{Shmakova,TT,TT2}
\begin{equation}
S_{BH}=\pi |p^{0}q_{0}|.
\end{equation}
Thus, one can conclude that the in the presence of a FI potential in the $%
SO\left( 2,2\right) $ covariant symplectic basis (of the $stu$
model), the $D0$-$D6$ BH charge configuration does not support any
critical point of the effective potential
$\tilde{V}_{eff,red,SO\left( 2,2\right) }$, and thus no attractor
solutions are admitted when only $p^{0}$ and $q_{0}$ are
non-vanishing (in the $\left( SO(1,1)\right) ^{2}$ covariant
symplectic basis).

\subsubsection{\label{lundi5}$\left( SO(1,1)\right) ^{2}$}

As it can be seen from Eqs. (\ref{scalpotstu2}) and
(\ref{scalpothet2}), in the case $\xi _{1}=0$ one obtains that
$V_{FI,\left( SO(1,1)\right) ^{2}}\varpropto V_{FI,SO\left(
2,2\right) }$, and everything, up to some redefinitions, goes as in
the $SO\left( 2,2\right) $ symplectic frame.

As done for the treatment of magnetic BH charge configuration, we
will make the simplifying assumption (\ref{lunnn1}), which we recall
to correspond to the most symmetric implementation of the FI
potential in the considered symplectic frame. Under such an
assumption, one gets
\begin{equation}
\tilde{V}_{eff,red,\left( SO(1,1)\right) ^{2}}=\frac{\sqrt{x^{2}y^{2}z^{2}+%
\frac{g^{2}}{3}(xy+xz+xy)(1+x^{2}y^{2}z^{2})}-xyz}{\frac{g^{2}}{3}(xy+xz+xy)}%
.
\end{equation}
Attractor Eqs. for such a potential yield the following solutions:
\begin{gather}
\frac{\partial \tilde{V}_{eff,red,\left( SO(1,1)\right) ^{2}}}{\partial x}=%
\frac{\partial \tilde{V}_{eff,red,\left( SO(1,1)\right) ^{2}}}{\partial y}=%
\frac{\partial \tilde{V}_{eff,red,\left( SO(1,1)\right) ^{2}}}{\partial z}=0;
\notag \\
\Updownarrow  \notag \\
\left\{
\begin{array}{l}
x=y=z; \\
\\
3x^{4}(-1+x^{6})+g^{2}(-1+2x^{6})^{2}=0.
\end{array}
\right.  \label{lun-1}
\end{gather}
Differently from the case of magnetic BH charge configuration (given by Eq. (%
\ref{tragedia1})), the algebraic equation of degree $12$ on the right-hand
side of Eq. (\ref{lun-1}) does not have an analytical solution. The
corresponding BH entropy can be computed in a parametric way as follows:
\begin{equation}
S_{BH,\left( SO(1,1)\right) ^{2}}=\pi \left| p^{0}q_{0}\right| \frac{\left(
2x^{6}-1\right) }{x^{3}},  \label{very-bad-1}
\end{equation}
yielding the restriction $\frac{1}{2}<x^{6}\leqslant 1$ on the solutions of
the algebraic equation of degree $12$ on the right-hand side of Eq. (\ref
{lun-1}).

\subsubsection{\label{lundi6}$SO\left( 8\right) $-truncated}

As it can be seen from Eqs. (\ref{scalpotmor2}) and (\ref{scalpothet2}), in
the case $\tilde{\xi}_{2}=\tilde{\xi}_{3}=0$ one obtains that $%
V_{FI,SO\left( 8\right) }\varpropto V_{FI,SO\left( 2,2\right) }$,
and everything, up to some redefinitions, goes as in the $SO\left(
2,2\right) $ symplectic frame.

As done for the treatment of magnetic BH charge configuration, we
will make the simplifying assumption (\ref{monday-1}), which we
recall to correspond to the most symmetric implementation of the
\textit{FI gauging} in the considered symplectic frame. Under such
an assumption, one gets
\begin{equation}
\tilde{V}_{eff,red,SO\left( 8\right) }=\frac{\sqrt{1+\frac{\tilde{\alpha}%
_{SO\left( 8\right) }}{x^{3}}(1+x^{6})(\frac{1}{x}+x)}-1}{\tilde{\alpha}%
_{SO\left( 8\right) }(1/x+x)},
\end{equation}
where (without loss of generality, under the assumption (\ref{monday-1})) we
put $x=y=z$.

Attractor Eq. for such a potential yields the following $\tilde{\alpha}%
_{SO\left( 8\right) }$-independent solution:
\begin{equation}
\frac{\partial \tilde{V}_{eff,red,SO\left( 8\right) }}{\partial x}%
=0\Leftrightarrow x=1,  \label{mmale!1}
\end{equation}
which is formally the same solution obtained for the magnetic BH charge
confoguration. The corresponding BH entropy reads
\begin{equation}
S_{BH,SO\left( 8\right) }=\pi |q_{0}p^{0}|\frac{\sqrt{1+4\tilde{\alpha}%
_{SO\left( 8\right) }}-1}{2\tilde{\alpha}_{SO\left( 8\right) }}.
\end{equation}
Consistently, in the ungauged limit $\tilde{\alpha}_{SO\left( 8\right)
}\rightarrow 0$ the well-known expression of the BH entropy is recovered
\cite{Shmakova,TT,TT2}:
\begin{equation}
lim_{g\rightarrow 0}S_{BH,SO\left( 8\right) }=\pi |q_{0}p^{0}|,
\end{equation}
whereas $S_{BH,SO\left( 8\right) }$ vanishes in the limit $\tilde{\alpha}%
_{SO\left( 8\right) }\rightarrow \infty $.

As it can be seen by recalling Eqs. (\ref{lun2})-(\ref{lun3}), the critical
points of $\tilde{V}_{eff,red,SO\left( 8\right) }$ necessarily have at least
one matter charge vanishing, depending on the sign of charges $p^{0}$ and $%
q_{0}$.

\section{\label{Stability}Analysis of Stability}

In order to determine the actual stability of the critical points of $%
V_{eff} $ found in the previous Section for the magnetic and $D0-D6$
BH charge configurations in various symplectic bases, one has to
study the sign of the eigenvalues of the (real form of the) Hessian
of $V_{eff}$ case
by case. It is worth pointing out that this is possible only in the \textit{%
effective potential formalism} studied here, which is intrinsically \textit{%
off-shell }(indeed, the treatment of Sects. \ref{General} and \ref
{General-stu} holds in the whole scalar manifold; only when
computing the critical points of $V_{eff}$ in the various symplectic
frames, as done in
Sect. \ref{BH-Entropies}, one goes \textit{on-shell}). The so-called \textit{%
entropy function formalism} (see \textit{e.g.} \cite
{Sen-old1,Sen-old2,Ebra1,Ebra2,Morales,Gao,Sen-review,ANYY1}) is
intrinsically \textit{on-shell}, and it does not allow one to study
analytically the stability of the attractor scalar configurations.

The key feature is the fact that in the (three different symplectic
frames of the) $stu$ model $V_{FI}$ is independent of the axions $%
(x_{1},y_{1},z_{1})$.

By denoting $f\equiv \{x,y,z\}$, the first derivatives of $V_{eff}$ with
respect to the axions $\partial _{f_{1}}V_{eff}$ read
\begin{equation}
\frac{\partial V_{eff}}{\partial f{_{1}}}=\frac{\partial V_{eff}}{\partial
V_{BH}}\frac{\partial V_{BH}}{\partial f{_{1}}}.
\end{equation}
By recalling Eq. (\ref{Veff-def}) one gets that $\frac{\partial V_{eff}}{%
\partial V_{BH}}=\frac{1}{\sqrt{1-4VV_{BH}}}$, which is strictly positive in
the region of definition $V_{BH}V<\frac{1}{4}$ of $V_{eff}$; consequently,
as already pointed out above:
\begin{equation}
\frac{\partial V_{eff}}{\partial f{_{1}}}=0\quad \Leftrightarrow \quad \frac{%
\partial V_{BH}}{\partial f{_{1}}}=0.  \label{axhoriz}
\end{equation}
In the $stu$ model such criticality conditions along the axionic directions
imply, both in the magnetic and $D0-D6$ BH charge configurations, $%
x_{1}=y_{1}=z_{1}=0$ (see \textit{e.g.} \cite{Ceresole}), \textit{i.e.} the
vanishing of all axions.

Let us now investigate the second order derivatives of $V_{eff}$ at its
critical points. Using the axionic criticality conditions (\ref{axhoriz})
one can easily show that ($g\equiv \{x,y,z\}$)
\begin{eqnarray}
\left. \frac{\partial ^{2}V_{eff}}{\partial f_{1}\partial g_{1}}\right| _{%
\frac{\partial V_{eff}}{\partial f{_{1}}}=0} &=&\left. \frac{\partial V_{eff}%
}{\partial V_{BH}}\right| _{\frac{\partial V_{eff}}{\partial f{_{1}}}%
=0}\left. \frac{\partial ^{2}V_{BH}}{\partial f_{1}\partial g_{1}}\right| _{%
\frac{\partial V_{eff}}{\partial f{_{1}}}=0}, \\
&&  \notag \\
\left. \frac{\partial ^{2}V_{eff}}{\partial f_{1}\partial g_{2}}\right| _{%
\frac{\partial V_{eff}}{\partial f{_{1}}}=0} &=&\left. \frac{\partial V_{eff}%
}{\partial V_{BH}}\right| _{\frac{\partial V_{eff}}{\partial f{_{1}}}%
=0}\left. \frac{\partial ^{2}V_{BH}}{\partial f_{1}\partial g_{2}}\right| _{%
\frac{\partial V_{eff}}{\partial f{_{1}}}=0}.
\end{eqnarray}
By straightforward calculations, one can show that all axionic-dilatonic
mixed derivatives vanish at $\frac{\partial V_{eff}}{\partial f{_{1}}}=0$:
\begin{equation}
\left. \frac{\partial ^{2}V_{BH}}{\partial f_{1}\partial g_{2}}\right| _{%
\frac{\partial V_{eff}}{\partial f{_{1}}}=0}=0.
\end{equation}
This implies that the $6\times 6$ (real form of the ) critical Hessian $%
\left( \frac{\partial ^{2}V_{eff}}{\partial f_{i}\partial g_{j}}\right) _{%
\frac{\partial V_{eff}}{\partial f_{k}}=0}$ ($k=1,2$) is block-diagonal. The
first $3\times 3$ block is the axionic one $\left. \frac{\partial ^{2}V_{eff}%
}{\partial f_{1}\partial g_{1}}\right| _{\frac{\partial V_{eff}}{\partial
f_{k}}=0}$, whereas the second is the $3\times 3$ dilatonic one $\left.
\frac{\partial ^{2}V_{eff}}{\partial f_{2}\partial g_{2}}\right| _{\frac{%
\partial V_{eff}}{\partial f_{k}}=0}$. Clearly, the real eigenvalues of such
two blocks can be investigated separately.

It is worth pointing out that in the dilatonic block $\left. \frac{\partial
^{2}V_{eff}}{\partial f_{2}\partial g_{2}}\right| _{\frac{\partial V_{eff}}{%
\partial f_{k}}=0}$ the axions $x_{1},y_{1},z_{1}$ are like parameters, and
one can implement the criticality condition $x_{1}=y_{1}=z_{1}=0$ \textit{%
before }differentiating with respect to dilatons without losing any
generality. This means that one can use the reduced form $V_{eff,red}$ of
the effective potential in order to study the sign of the eigenvalues of the
dilatonic block of the critical Hessian of $V_{eff}$, which we will denote
by $\zeta _{1}$, $\zeta _{2}$ and $\zeta _{3}$.

Concerning the axionic block $\left. \frac{\partial ^{2}V_{eff}}{\partial
f_{1}\partial g_{1}}\right| _{\frac{\partial V_{eff}}{\partial f_{k}}=0}$,
in the $D0-D6$ BH charge configuration it has strictly positive eigenvalues
(as it can be shown by straightforward calculations by recalling that $%
x_{2}y_{2}z_{2}>0$ - see Eq. (\ref{CERNN1}) - ). On the other hand, in the
magnetic BH charge configuration the axionic block is proportional (through
a strictly positive coefficient) to the following $3\times 3$ matrix
\begin{equation}
\left(
\begin{array}{ccc}
\begin{array}{c}
\left| \frac{q_{0}p^{2}p^{3}}{p^{1}}\right| (y^{2}+z^{2}) \\
~
\end{array}
&
\begin{array}{c}
p^{3}q_{0}\left[ \aleph z^{2}-1\right] \\
~
\end{array}
&
\begin{array}{c}
p^{2}q_{0}\left[ \aleph y^{2}-1\right] \\
~
\end{array}
\\
\begin{array}{c}
p^{3}q_{0}\left[ \aleph z^{2}-1\right] \\
~
\end{array}
&
\begin{array}{c}
\left| \frac{q_{0}p^{1}p^{3}}{p^{2}}\right| (x^{2}+z^{2}) \\
~
\end{array}
&
\begin{array}{c}
p^{1}q_{0}\left[ \aleph x^{2}-1\right] \\
~
\end{array}
\\
p^{2}q_{0}\left[ \aleph y^{2}-1\right] & p^{1}q_{0}\left[ \aleph x^{2}-1%
\right] & \left| \frac{q_{0}p^{1}p^{2}}{p^{3}}\right| (x^{2}+y^{2}).
\end{array}
\right) _{\frac{\partial V_{eff}}{\partial f_{2}}=0},
\end{equation}
where $\aleph \equiv sgn\left( q_{0}p^{1}p^{2}p^{3}\right) $, and the
rescaled dilatons are defined by Eq. (\ref{male-male}). In the following we
will denote the eigenvalues of the axionic block of the critical Hessian of $%
V_{eff}$ by $\chi _{1}$, $\chi _{2}$ and $\chi _{3}$.

Let us briefly recall the stability of critical points of $V_{BH}$ in the
\textit{ungauged} $stu$ model. The $\frac{1}{2}$-BPS and non-BPS $Z=0$
critical points are actually attractors in the strict sense, with no
vanishing eigenvalues of the Hessian at all: $\chi _{1},\chi _{2},\chi
_{3},\zeta _{1},\zeta _{2},\zeta _{3}>0$. On the other hand, the non-BPS $%
Z\neq 0$ critical points are stable, with two flat directions permaining at
all orders, and spanning the moduli space $\left( SO\left( 1,1\right)
\right) ^{2}$ (\textit{i.e.} the scalar manifold of the \textit{ungauged} $%
stu$ model in $d=5$; see \cite{TT,BFGM1,TT2,fm07,Ceresole}): one gets $\chi
_{1}>0,\chi _{2}=\chi _{3}=0$, $\zeta _{1},\zeta _{2},\zeta _{3}>0$ and $%
\chi _{1},\chi _{2},\chi _{3}>0$, $\zeta _{1}>0,\zeta _{2}=\zeta _{3}=0$,
respectively for the magnetic and $D0-D6$ \ BH charge configurations. It is
also worth recalling that the magnetic BH charge configuration supports in
general all three classes of critical points of $V_{BH}$, whereas the $D0-D6$
BH charge configuration supports only non-BPS $Z\neq 0$ critical points of $%
V_{BH}$.

In the following we will report the results concerning the stability
of the critical points of $V_{eff}$ in the various considered
symplectic frames of the $stu$ model with FI terms, both in the
magnetic and in the $D0-D6$ \ BH charge configurations.

\subsection{\label{Magnetic2}Magnetic Configuration}

As given by Eq. (\ref{male!-1}), in the $SO\left( 2,2\right) $
covariant symplectic frame two sets of critical points of
$\tilde{V}_{eff,red,SO\left( 2,2\right) }$ exist. For both solutions
one gets the following result:
\begin{equation}
\text{\textit{$SO\left( 2,2\right) $ sympl. frame:~}}\left\{
\begin{array}{l}
q_{0}p^{1}p^{2}p^{3}>0:\left\{
\begin{array}{l}
\chi _{1},\chi _{2},\chi _{3}>0; \\
\\
\zeta _{1},\zeta _{2},\zeta _{3}>0;
\end{array}
\right. \\
\\
q_{0}p^{1}p^{2}p^{3}<0:\left\{
\begin{array}{l}
\chi _{1},\chi _{2}>0,\chi _{3}=0~~\left( \tilde{\alpha}=0:\chi _{1}>0,\chi
_{2}=\chi _{3}=0\right) ; \\
\\
\zeta _{1},\zeta _{2},\zeta _{3}>0.
\end{array}
\right.
\end{array}
\right.  \label{dep1}
\end{equation}

On the other hand, as given by Eq. (\ref{tragedia1}), in the $\left(
SO(1,1)\right) ^{2}$ covariant symplectic basis in the simplifying
assumption (\ref {lunnn1}) only one set of critical points of
$\tilde{V}_{eff,red,\left( SO(1,1)\right) ^{2}}$ exist, for which it
holds that
\begin{equation}
\mathit{\left( SO(1,1)\right) ^{2}}\text{\textit{\ sympl.
frame:~}}\left\{
\begin{array}{l}
q_{0}p^{1}p^{2}p^{3}>0:\left\{
\begin{array}{l}
\chi _{1},\chi _{2},\chi _{3}>0; \\
\\
\zeta _{1},\zeta _{2},\zeta _{3}>0;
\end{array}
\right. \\
\\
q_{0}p^{1}p^{2}p^{3}<0:\left\{
\begin{array}{l}
\chi _{1}>0,\chi _{2}<0,\chi _{3}<0,~ \\
\left( g=0:\chi _{1}>0,\chi _{2}=\chi _{3}=0\right) ; \\
\\
\zeta _{1},\zeta _{2},\zeta _{3}>0.
\end{array}
\right.
\end{array}
\right.  \label{dep2}
\end{equation}

Finally, as given by Eq. (\ref{male!2}), in the $SO\left( 8\right) $%
-truncated symplectic basis in the simplifying assumption
(\ref{monday-1}) only one set of critical points of
$\tilde{V}_{eff,red,SO\left( 8\right) }$ exist, and it can be
computed that
\begin{equation}
\text{\textit{$SO\left( 8\right) $-trunc. sympl. frame:~}}\left\{
\begin{array}{l}
q_{0}p^{1}p^{2}p^{3}>0:\left\{
\begin{array}{l}
\chi _{1},\chi _{2},\chi _{3}>0; \\
\\
\zeta _{1},\zeta _{2},\zeta _{3}>0;
\end{array}
\right. \\
\\
q_{0}p^{1}p^{2}p^{3}<0:\left\{
\begin{array}{l}
\chi _{1}>0,\chi _{2}=\chi _{3}=0; \\
\\
\zeta _{1},\zeta _{2},\zeta _{3}>0.
\end{array}
\right.
\end{array}
\right.  \label{dep3}
\end{equation}

Let us briefly analyze the obtained results
(\ref{dep1})-(\ref{dep2}). As intuitively expected, the introduction
of the (FI) potential never affects the sign of those eigenvalues
strictly positive in the ungauged limit, and thus
it does not modify the stability of those gauged solutions (supported by $%
q_{0}p^{1}p^{2}p^{3}>0$) which in the ungauged limit are
$\frac{1}{2}$-BPS or non-BPS $Z=0$. On the other hand, by looking at
those gauged solutions (supported by $q_{0}p^{1}p^{2}p^{3}<0$) which
in the ungauged limit are non-BPS $Z\neq 0$, one realizes that in
the presence of (FI) potential the possible lift of the two ungauged
non-BPS $Z\neq 0$ flat directions strictly depends on the symplectic
basis being considered. Indeed, such flat directions both persist in
the $SO\left( 8\right) $-truncated frame, only one is lifted up to a
positive direction in the $SO\left( 2,2\right) $ covariant basis,
and both are lifted down to tachionic (\textit{i.e.} negative)
directions in the $\left( SO(1,1)\right) ^{2}$ covariant basis.

\subsection{\label{D0-D6-2}$D0$-$D6$ Configuration}

As given by Eqs. (\ref{lunn1}) and (\ref{lunn2}), the
$SO\left(2,2\right) $ covariant symplectic basis does not admit any
gauged ($\widetilde{\alpha }\neq 0$) critical point of
$\tilde{V}_{eff,red,SO\left( 2,2\right) }$ in the $D0$-$D6$ BH
charge configuration.

On the other hand, as given by Eqs. (\ref{lun-1}) and (\ref{very-bad-1}),
despite the fact that analytical critical points of $\tilde{V}%
_{eff,red,\left( SO(1,1)\right) ^{2}}$ cannot be computed, in the
$\left( SO(1,1)\right) ^{2}$ symplectic frame one can still study
the sign of the eigenvalues of the critical Hessian of
$\tilde{V}_{eff,red,\left( SO(1,1)\right) ^{2}}$, thus obtaining
\begin{equation}
\mathit{\left( SO(1,1)\right) ^{2}}\text{\textit{\ cov. sympl. frame:~}}%
\left\{
\begin{array}{l}
\chi _{1},\chi _{2},\chi _{3}>0; \\
\\
\zeta _{1}>0,\zeta _{2}>0,\zeta _{3}>0~~\left( g=0:\zeta _{1}>0,\zeta
_{2}=\zeta _{3}=0\right) .
\end{array}
\right.  \label{dep4}
\end{equation}

Finally, as given by Eq. (\ref{mmale!1}), in the $SO\left( 8\right) $%
-truncated symplectic basis in the simplifying assumption
(\ref{monday-1}) only one set of critical points of
$\tilde{V}_{eff,red,SO\left( 8\right) }$ exist, and it can be
computed that
\begin{equation}
\text{\textit{$SO\left( 8\right) $-trunc. sympl. frame:~}}\left\{
\begin{array}{l}
\chi _{1},\chi _{2},\chi _{3}>0; \\
\\
\zeta _{1}>0,\zeta _{2}<0,\zeta _{3}<0~~\left( \tilde{\alpha}_{SO\left(
8\right) }=0:\zeta _{1}>0,\zeta _{2}=\zeta _{3}=0\right) .
\end{array}
\right.  \label{dep5}
\end{equation}

Once again, let us briefly analyze the obtained results
(\ref{dep4})-(\ref {dep5}). As intuitively expected, also in the
$D0-D6$ BH charge configuration the introduction of the (FI)
potential never affects the sign of those eigenvalues strictly
positive in the ungauged limit. Furthermore, by looking at Eqs.
(\ref{dep4})-(\ref{dep5}) one can realizes once again that in the
presence of (FI) potential the possible lift of the two ungauged
non-BPS $Z\neq 0$ flat directions strictly depends on the symplectic
frame being considered. Indeed, such flat directions are both lifted
up to positive directions in the $\left( SO(1,1)\right) ^{2}$
covariant basis, whereas they
are both lifted down to tachionic directions in the $SO\left( 8\right) $%
-truncated frame (as it was the case for $\chi _{2}$ and $\chi _{3}$ in the $%
\left( SO(1,1)\right) ^{2}$ basis for the magnetic BH charge
configuration).

\section{\label{Conclusion}Further Developments}

It is clear that the present study paves the way to a number of possible
further developments. One of particular relevance is the following.

By considering the issue of stability in the magnetic BH charge
configuration, one might wonder whether $\chi _{3}=0$ in the
$SO\left( 2,2\right) $ frame (see Eq. (\ref{dep1})) and $\chi
_{2}=\chi _{3}=0$ in the $SO\left( 8\right) $-truncated frame (see
Eq. (\ref{dep3})) actually are
\textit{flat directions} of the corresponding effective potentials, \textit{%
i.e.} whether they persist also at higher order in the covariant
differentiation of such (not symplectic-invariant) potentials. Since in
general the geometry of the scalar manifold is not affected by the
introduction of the gauging, in case of positive answer it would be
interesting to determine the moduli space spanned by such \textit{flat
directions}. It is clear that the reasoning about \textit{flat directions}
performed in \cite{fm07} should not apply in the considered cases, because
in the presence of (FI, but likely for a completely general) gauging $%
V_{eff} $\ is not symplectic-invariant, as instead $V_{BH}$\ is by
definition. We leave this issue for future work.

\section*{\textbf{Acknowledgments}}

We would like to warmly acknowledge L. Andrianopoli, R. D'Auria, P. Fr\'{e}
and M. Trigiante for enlighting discussions and interest in this work.

A. M. would also like to thank the Department of Physics, Theory Unit Group
at CERN, where part of this work was done, for kind hospitality and
stimulating environment.

The work of S.B. has been supported in part by the European Community Human
Potential Program under contract MRTN-CT-2004-005104 \textit{``Constituents,
fundamental forces and symmetries of the Universe''}.

The work of S.F.~has been supported in part by the European Community Human
Potential Program under contract MRTN-CT-2004-005104 \textit{``Constituents,
fundamental forces and symmetries of the Universe''}, in association with
INFN Frascati National Laboratories and by D.O.E.~grant DE-FG03-91ER40662,
Task C.

The work of A.M. has been supported by a Junior Grant of the \textit{%
``Enrico Fermi''} Centre, Rome, in association with INFN Frascati National
Laboratories.

The work of A.Y. was supported in part by the grant INTAS-05-7928, in
association with INFN Frascati National Laboratories. 

\end{document}